\documentclass[11pt]{article} 
\usepackage{mystyle-new}
\usepackage{authblk}
\usepackage[T1]{fontenc}
\usepackage{epsfig,amsmath} 
\usepackage[colorlinks=true, allcolors=blue]{hyperref}
\usepackage{color}
\usepackage{graphicx}
\usepackage{orcidlink}
\usepackage{cprotect}
\usepackage{xspace}
\definecolor{red}{rgb}{1,0,0}
\def\lesssim{\ \hbox{\raise 2pt \hbox{$<$} \kern -13pt
                     \lower 3pt \hbox{$\sim$}}\ }
\def\greatersim{\ \hbox{\raise 2pt \hbox{$>$} \kern -13pt
                     \lower 3pt \hbox{$\sim$}}\ }

\def\lsim{\mathrel{\rlap{\lower4pt\hbox{\hskip1pt$\sim$}}
    \raise1pt\hbox{$<$}}}                
\def\gsim{\mathrel{\rlap{\lower4pt\hbox{\hskip1pt$\sim$}}
    \raise1pt\hbox{$>$}}}                

\def\cascade{{\scshape Cascade}3\xspace}
\def\pythia{{\scshape Pythia}8\xspace}
\def\pythiaPB{{\scshape{Pdf2Isr}}\xspace}
\def\herwig{{\scshape Herwig}\xspace}
\def\sherpa{{\scshape Sherpa}\xspace}
\def\powheg{{POWHEG}\xspace}

\input epsf.tex
\def\desepsf(#1 width #2){\epsfxsize=#2 \epsfbox{#1}}
\def\kt{\ensuremath{k_{{\rm t}}}}
\def\kti#1{\ensuremath{k_{{\rm t},#1}}}

\def\pt{\ensuremath{p_{\rm t}}}
\def\pti#1{\ensuremath{p_{t, #1}}}
\def\PZ{\ensuremath{Z}}

\def\qt{\ensuremath{q_{\rm t}}}
\def\qti#1{\ensuremath{q_{t, #1}}}
\def\zdyn{\ensuremath{z_{\rm dyn}}}
\def\zM{\ensuremath{z_{\rm M}}}

\newcommand{\sqrts}{\ensuremath{\sqrt s}}

\newcommand{\alphas}{\ensuremath{\alpha_\mathrm{s}}}

\newcommand{\asmz}{\ensuremath{\alphas(m_{\PZ})}}

\newcommand{\PBM}{PB\xspace}
\newcommand{\PBNLOset}{{PB-NLO-2018}\xspace}
\newcommand{\PBset}{{PB-LO-2018}\xspace}
\newcommand{\PBToyset}{{PB-Toy}\xspace}
\newcommand{\TMDlib}{{\scshape TMDlib}\xspace}
\newcommand{\TMDplotter}{{\scshape TMDplotter}\xspace}
\newcommand{\qcdnum}{{\scshape QCDnum}\xspace}

\newcommand{\MCatNLO}{{\sc MadGraph5\_aMC@NLO}\xspace}

\newcommand{\pstmd}{{\scshape Ps2Tmd}\xspace}

\newcommand{\as}{\ensuremath{\alpha_s}}

\newcommand{\GeV}{\text{GeV}}

\newenvironment{tolerant}[1]{\par\tolerance=#1\relax}{ \par }
\usepackage{amsmath,bm}
\usepackage{lineno}

\usepackage{cite,./mcite}
\usepackage{tikz}
\usepackage[symbol*]{footmisc}

\newcommand{\dglap}{Gribov:1972ri,Lipatov:1974qm,Altarelli:1977zs,Dokshitzer:1977sg}

\providecommand{\DOI}[1]{\href{http://dx.doi.org/#1}}

\def\updfevolv{{\sc uPDFevolv2}}

\begin{document} 
 
\title{A parton shower consistent with parton densities at LO and NLO: \pythiaPB }

\author[1,2]{H.~Jung\thanks{hannes.jung@desy.de}\orcidlink{0000-0002-2964-9845}}
\affil[1]{Deutsches Elektronen-Synchrotron DESY, Germany}
\affil[2]{II. Institut f\"ur Theoretische Physik, Universit\"at Hamburg,  Hamburg, Germany}
\author[3]{L.~L\"onnblad\thanks{leif.lonnblad@fysik.lu.se}\orcidlink{0000-0003-1269-1649}}
\affil[3]{Department of Physics, Lund University, Lund, Sweden}
\author[1]{M.~Mendizabal\thanks{mikel.mendizabal.morentin@desy.de}\orcidlink{0000-0002-6506-5177}}
\author[1]{S.~Taheri~Monfared\thanks{sara.taheri.monfared@desy.de}\orcidlink{0000-0003-2988-7859}}

\begin{titlepage} 
\maketitle
\vspace*{-10cm}
\begin{flushright}
DESY-25-062\\
MCNET-25-06
\end{flushright}
\vspace*{+7cm}
\end{titlepage}

\begin{abstract}

  We present a method for obtaining an initial-state parton shower
  model where the (backward) evolution fully consistent with the
  (forward) evolution of the collinear parton density used.

  As a proof-of-concept we use parton densities obtained with the
  Parton Branching (PB) approach, and modify the default initial-state
  shower in \pythia with this method to be consistent with them. PB is
  ideally suited for checking the validity of our method since, in
  addition to producing collinear parton densities, it also
  produces the corresponding transverse-dependent (TMD) ones, and these
  can then be directly compared to the transverse momentum
  distribution obtained from the parton shower.

  We show that TMD distributions which we in this way obtain from our
  modified \pythia shower using leading order (LO) parton densities
  and splitting functions are fully consistent with the corresponding
  leading order TMD densities. At next-to-leading order (NLO) it is
  not possible to achieve the same consistency using the built-in LO
  splitting functions in the shower, but we show that by introducing
  NLO splitting functions using a reweighting procedure, we can
  achieve consistency also at NLO.

  The method presented here, which we have named \pythiaPB, can be
  easily extended to any collinear parton densities, as long as the
  exact conditions for the evolution are known. With the \pythiaPB\, method we obtain an initial-state parton shower which in principle
  has no free parameters, and is fully consistent with collinear parton
  densities at LO and NLO.
  
\end{abstract} 

\newpage
\section {Introduction}
The description of precise measurements of processes involving high transverse momentum jets as well as precision measurements of vector-boson production require rather sophisticated methods. 
Only in rare cases a description using fixed-order perturbative calculations is sufficient. In most cases, a simulation, including multiple partonic radiation and hadronization, as performed in multi-purpose Monte Carlo event generators (MCEG) like \herwig~\cite{Bahr:2008pv}, \pythia~\cite{Bierlich:2022pfr}, and \sherpa~\cite{Bothmann:2019yzt,Gleisberg:2008ta}, is required. 

\begin{tolerant}{9000}
 The hard, perturbative process can be calculated  externally via packages like \MCatNLO~\cite{Alwall:2014hca} or \powheg~\cite{Oleari:2010nx,Alioli:2010xd}  at leading-order (LO) or  next-to-leading order (NLO)
accuracy, and  can be supplemented with initial- and final-state parton showers, as well as with multi-parton interactions and hadronization. While quite some effort has been put into matching and merging of parton showers with the NLO matrix element calculations~\cite{Frixione:2003ei,Frixione:2002ik,Frixione:2007vw,Nason:2012pr,Alwall:2014hca,Frederix:2015eii}, parton showers still appear to lack a direct correspondence with the parton densities used in the calculation of the hard process as well as in the backward evolution. 
In Ref.~\cite{Frixione:2023ssx}, it is argued that collinear parton densities, as well as NLO hard scattering coefficients, must be recalculated in a scheme that corresponds to the one used in parton showers, pointing to an inconsistency in the present treatment.
\end{tolerant}

In this paper we describe a method, called \pythiaPB, to construct
the initial-state radiation (ISR) simulated as a parton shower to
follow exactly the evolution of the collinear parton density by using
the Parton-Branching (\PBM
)-method~\cite{Hautmann:2017fcj,Hautmann:2017xtx} as a test-case. The
\PBM -approach has been developed as a method to solve the evolution
equations iteratively, in order to provide collinear as well as
Transverse Momentum Dependent (TMD) parton densities, by simulating
each individual branching and including the appropriate kinematic
relations.  TMD parton densities are ideal for testing the consistency
of the evolution and the parton shower, since they can be obtained
from both.  The advantage of the PB method is that all the details of
each individual branching processes are known and can be studied.  The
\PBM -TMD distributions agree by construction with the collinear
distribution upon integrating over the transverse momentum exactly.
In order to obtain  \PBM-TMD distributions  the evolution scale is interpreted as a physical scale with a relation to the transverse momentum of the emitted parton. 

In this study, we modify the default\footnote{There are several
  showers implemented in \pythia, the default one is called
  \texttt{SimpleSpaceShower}.} initial-state parton shower in \pythia
to use the parameters of the collinear \PBM\ parton distribution and
to follow the same kinematic constraints as in the parton evolution to
obtain effective TMD distributions.  We find that only minor
modifications of the \pythia code are needed to obtain TMD
distributions that are in perfect agreement with those from \PBM at LO. This illustrates and proves that the same physical picture is being used.
Going to NLO collinear parton densities, we show
that the use of LO splitting functions leads to inconsistent results,
and the implementation of NLO splitting functions into the initial-state radiation framework is required. We apply a method, described in
Refs.~\cite{Mrenna:2016sih,Lonnblad:2012hz} to properly treat negative contributions of NLO splitting functions at large $z$ (and
small \kt ) within a parton shower framework.

In the following, we briefly describe a method to obtain TMD distributions from general parton shower event generators. We will then apply this method to compare the TMD distributions obtained from \PBM\ with those from \pythia . We will then describe how the parton shower in \pythia\ can be modified to follow the same conditions as those used in \PBM -method at LO. We discuss in detail the use of collinear  parton densities obtained at NLO and show the importance of applying NLO splitting functions, as well as the same evolution method for \as\ at NLO. 
We comment on the frame dependence in the calculation of the transverse momentum\kt\ in the TMDs.

\section{TMDs from parton showers: \pstmd -method \label{pstmd_approach}}

\begin{figure} [t]
\centering
\includegraphics[width=0.4\linewidth]{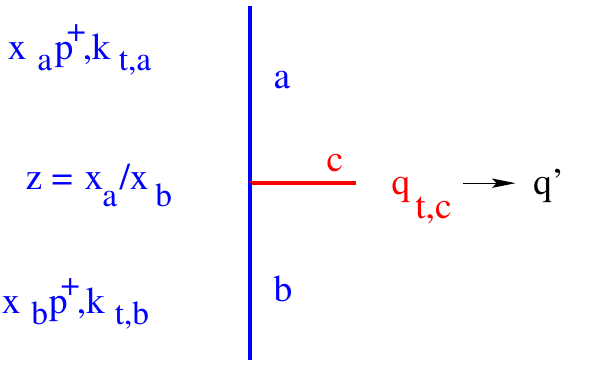} 
\caption {Typical parton branching process $b \to a +c$.}
\label{fig:kine}
\end{figure}

In a parton shower approach, each individual branching process is simulated using appropriate kinematics (for the notation see Fig.~\ref{fig:kine}). 
Due to the kinematic relations in each splitting, a transverse
momentum of the emitted parton $c$ as well as of the partons $a$ and
$b$ will appear.  After the full initial-state shower is generated, an effective final transverse momentum distribution can be reconstructed.
\begin{figure} [t]
\centering
\includegraphics[width=0.3\linewidth]{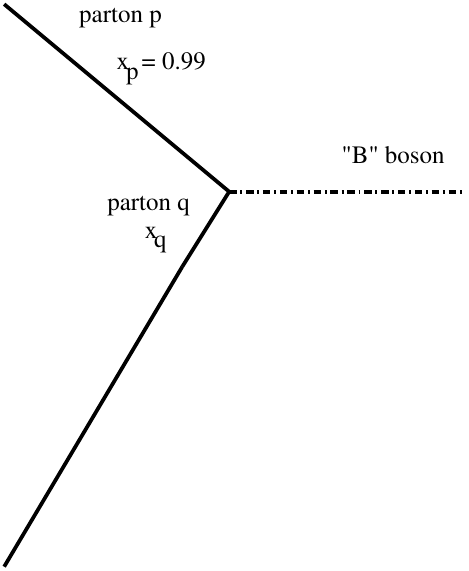}\hskip 2cm \includegraphics[width=0.3\linewidth]{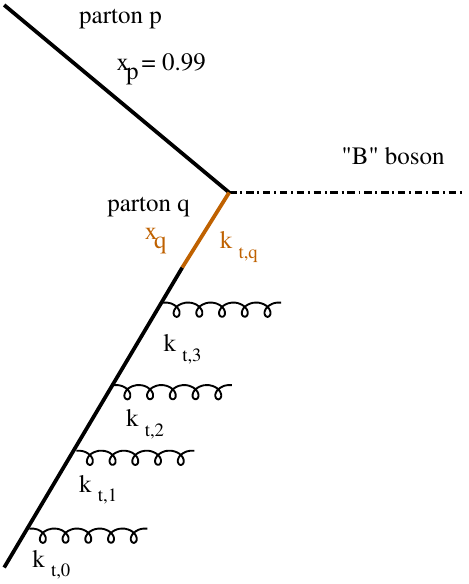}
\caption {Illustration of the toy process: $p + q \to B$: (left) bare process; (right) including initial-state parton shower}
\label{fig:Toy}
\end{figure}
This effective TMD distribution can be obtained from any parton shower MCEG with the \pstmd -method~\cite{Jung:2021jno,Schmitz:427383} (see
Fig.~\ref{fig:Toy}): A toy $2 \to 1$ process ($p + q \to B$) is
generated, where one initial parton has momentum fraction 
$x_p= 0.99$ and does not develop any initial-state radiation, while the other parton has varying $x_q$ according to the collinear parton
density and can develop an initial-state parton shower.  The produced
toy colorless ``$B$-boson'' particle is used to calculate the
kinematics and for easy identification in the event record. The
$B$-boson can couple equally to gluons and quarks and is
therefore unphysical. The scale $\mu$ of the process is generated over a
large range.
The transverse momentum of the initial parton $q$ is easily obtained
from the kinematics of the process:
\begin{equation}
\kti{q} = \pti{B} \,
\end{equation}
with $\pti{B}$ being the transverse momentum of the particle $B$ (the transverse momentum of parton $p$ is negligible by construction).
The cross section of the toy process consists only of the momentum weighted collinear parton density, at the generated $x_q$ and the scale $\mu$. The generated events are passed to a Rivet plugin~\cite{Bierlich:2019rhm}, where the momentum fraction $x_q$, the scale $\mu$ as well as the transverse momentum $\kti{q}$ are extracted. Their values are then stored in a grid  to be directly used within \TMDlib~\cite{Abdulov:2021ivr} and visualized with \TMDplotter .

\begin{tolerant}{8000}
The \pstmd -method has already been validated using \cascade~\cite{Baranov:2021uol}  and applied to  \pythia\ and \herwig~\cite{Jung:2021jno,Schmitz:427383}. 
\end{tolerant}

\section {\label{sec2} {\bfseries\PBM -method} and {\bfseries{\pythia }}  parton shower}

We start with a short summary of the main features of the \PBM -method, followed by a description of the different ordering conditions and brief overview of the \pythia\ parton shower method. We then compare TMD distributions obtained with the \PBM -method to those from the \pythia\ initial-state parton shower.

\subsection{\PBM -method}
The DGLAP evolution equation~\cite{\dglap} for the momentum-weighted parton density \(xf_a(x,\mu^2)\) of parton \(a\) with momentum fraction \(x\) at scale \(\mu\) is written as:
\begin{equation}
\label{EvolEq}
 \mu^2 \frac{\partial (x f_a(x,\mu^2))}{\partial \mu^2} = \sum_b \int_x^{1} dz \; P_{ab} \left(z,\as\right) \; \frac{x}{z} f_b\left(\frac{x}{z}, \mu^2\right),
\end{equation}
where \(P_{ab}\) represents the regularized DGLAP splitting functions, describing the transition of parton \(b\) into parton \(a\).

After replacing the plus-prescription in $P_{ab}$ with a Sudakov form factor, $\Delta_a$,
the solution of  the evolution equation for momentum-weighted parton densities, \(xf_a(x,\mu^2)\), at scale \(\mu\) can be written as (e.g.~\cite{Ellis:1991qj}):
\begin{equation}
\label{sudintegral2}
x f_a(x,\mu^2) = \Delta_a(\mu^2) x f_a(x,\mu^2_0) + \sum_b \int_{\mu^2_0}^{\mu^2} \frac{dq^{\prime 2}}{q^{\prime 2}} \frac{\Delta_a(\mu^2)}{\Delta_a(q^{\prime 2})} \int_x^{\zM} dz \;  P^{(R)}_{ab}(z, \as) \frac{x}{z} f_b\left(\frac{x}{z}, q^{\prime 2}\right),
\end{equation}
where $P^{(R)}$ are the real, unregularized splitting functions\footnote{replacing $1/(1 - z)_+$  by $1/(1 - z)$ and without the virtual contribution}, 
 \(\mu_0\) is the starting scale, $\Delta_a(\mu^2) := \Delta_a(\mu^2, \mu_0^2)$ is the Sudakov form factor and  ${\bf q}^{\prime}$ is a 2-dimensional vector with ${\bf q}^{\prime \,2} = q^{\prime 2} $. 
From the comparison of eq.(\ref{EvolEq}) with eq.(\ref{sudintegral2}), one can immediately see that, for consistency, $\zM \to 1$. However, for numerical reasons, $\zM=1-\epsilon$ with very small $\epsilon$ to avoid the $1/(1-z)$ singularity in splitting functions.

The \PBM\ approach provides a method to solve the evolution equation
by an iterative method, applying the concept of Sudakov form factors,
as described in Refs.~\cite{Hautmann:2017fcj,Hautmann:2017xtx}. The
advantage of this iterative approach is that each individual splitting
process is simulated, allowing for proper treatment of the kinematic
relations of the splitting. This method has been applied to determine
collinear and TMD distributions by fitting the parameters of the
initial distribution~\cite{Martinez:2018jxt} such that deep-inelastic
measurements at HERA~\cite{Abramowicz:2015mha} can be well described
over a wide range in $x$ and $Q^2$.

Two different sets were obtained in Ref.~\cite{Martinez:2018jxt},
depending on the scale choice in \as : in \PBNLOset~Set1 the evolution
scale $ q' $ was used as the scale in
\as\ 
resulting in collinear distributions identical to those obtained as
HERAPDF; in \PBNLOset~Set2 the transverse momentum \qt\ (for a
definition see next section) was used as the scale in \as , and
different collinear and TMD distributions were obtained, with a
similar $\chi^2/ndf \sim 1.2$.
This scale choice for \as\ is motivated from angular ordering, and
leads to two different regions: a perturbative region, with
$\qt > q_{\rm cut}$, and a non-perturbative region of $\qt < q_{\rm cut}$, where \as\
is frozen at $q_0$.  The initial distributions were defined at a scale
$\mu_0=1.374 (1.181)$ \GeV\ for \PBNLOset~Set1(Set2).

\subsection{Ordering conditions in \PBM -method}
The DGLAP evolution equations allow the determination of the parton densities at a scale $\mu$ if they are  known at a different scale $\mu_0$. However, these equations do not provide a physical interpretation of the evolution scale. In parton shower approaches, as well as in the \PBM -method,  the DGLAP equations are extended by giving a physical interpretation to the evolution scale.  A typical branching process, $b \to a + c$, is shown in Fig.~\ref{fig:kine}, with the  light-cone momenta $ p^+_a  = z p^+_b$, $p^+_c = (1 - z ) p^+$ with $p^+$ being the light-cone momentum of the  beam particle.

The transverse momentum $\qti{c}$ can be calculated from the evolution scale $\mu$ in different ways:
\begin{itemize}\setlength{\itemsep}{0pt}
\item \pt -ordering: the transverse momentum $\qti{c}$ is directly associated with the evolution scale $\mu$, such that $\qti{c} = q^\prime $
\item angular ordering: the rescaled transverse momentum $\qti{c}/(1-z)$ is related to the polar angle $\Theta_c$ of the emitted parton, which is taken as the evolution scale, resulting in $\qti{c} =  (1 - z ) q^\prime $.
\end{itemize}

Ref.~\cite{Hautmann:2017xtx} presents transverse momentum distributions obtained from \pt - and angular ordering, illustrating significant differences between them.

\subsection{Initial-state shower in \pythia\  }
\label{sec:pythia-isr}

The initial-state parton shower in \pythia\ starts  from the hard scattering using a backward evolution applying ratios of collinear parton densities. 
A detailed description of the parton shower approach is given in Refs.~\cite{Bengtsson:1986gz,Bierlich:2022pfr}.

The probability for an emission is given by the Sudakov form factor for backward evolution:
\begin{equation}
\log \Delta_{bw} (z, \mu, \mu_{i-1}) = - \sum_b \int_{\mu_{i-1}^2}^{\mu^2}  \frac{d q^{\prime\,2}}{q^{\prime\,2}}\int_x^{\zdyn} dz P^{(R)}_{ab} (\as(z,q'),z) \frac{x' f_b(x',q')}{xf_a(x,q')} \; .
\label{Suda}
\end{equation}

By default, the scale $q'$ is the transverse momentum of the emitted parton. Additional corrections are applied for heavy flavoured partons. The default ordering condition in \pythia\ is transverse momentum (\pt-) ordering, implying the evolution scale $q' = \pt$. 
The integration limit \zdyn\ is constrained by the masses of the radiating dipole system, with $\zdyn < 1$. In general, \zdyn\ in eq.(\ref{Suda}) does not match \zM\ in eq.(\ref{sudintegral2}). 

It should be noted that the splitting probability by default is
smoothly suppressed for small transverse momenta by a factor
$p_{\perp}^2/(p_\perp^2 + p_{\perp0}^2)$, together with a hard cutoff
at $p_{\perp\rm min}$.

\subsection{PB-TMDs and effective TMDs from \pythia}
In the following we compare the predictions from \PBM\ Set1 with those obtained from \pythia\ initial-state parton shower using parameters from the  CUET tune\footnote{\pythia setting: {\tt Tune:pp = 18}.}.
We apply the \pstmd -method using  \pythia\  (version 8.311) to generate the initial-state  (space-like) parton shower. The distributions are obtained by running the toy process described in Section~\ref{pstmd_approach} at $\sqrts = 5 \cdot 10^6$~\GeV\ to ensure appropriate coverage of the phase space.
\begin{figure} [t]
\centering
\includegraphics[width=0.35\linewidth]{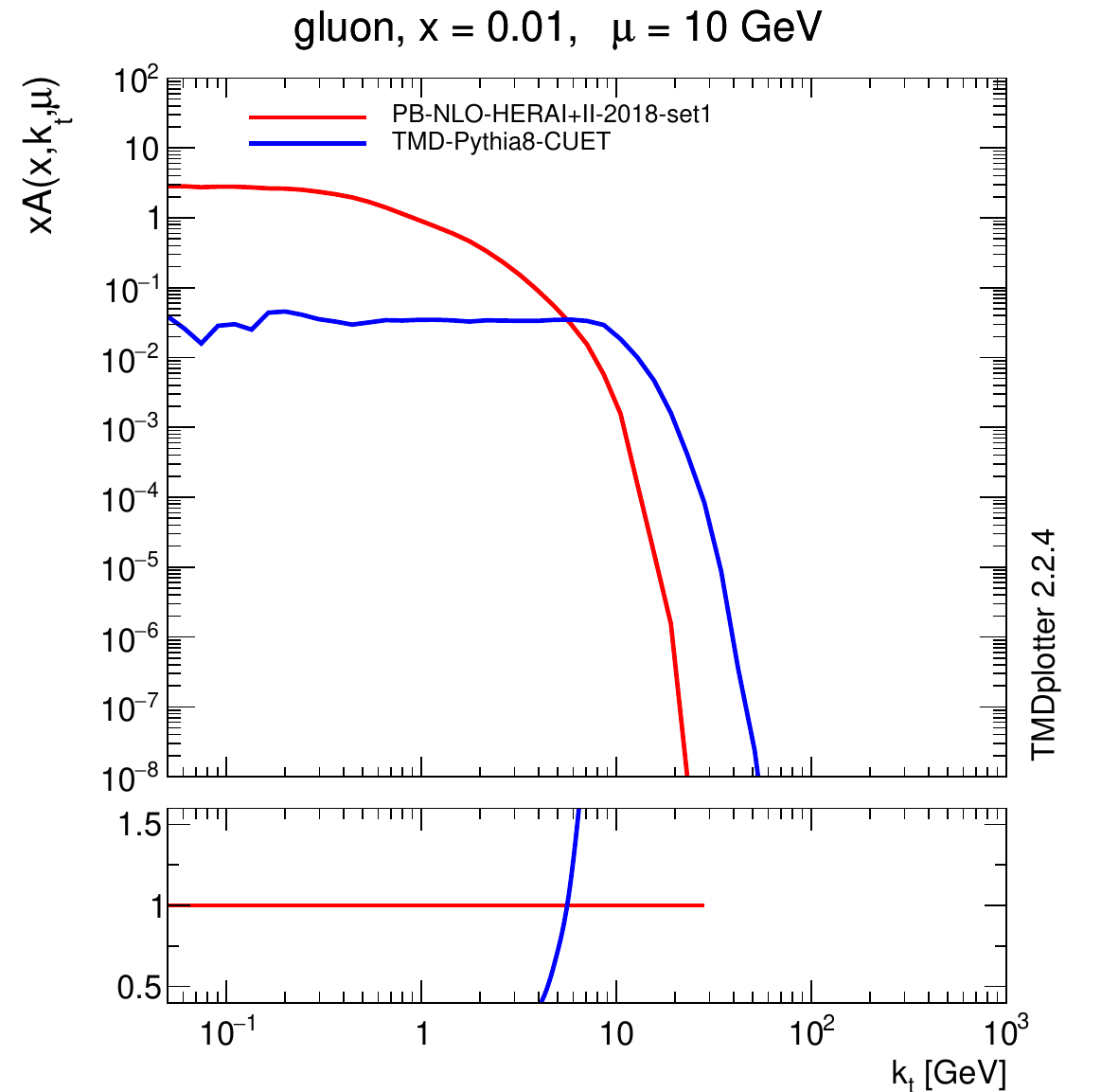} 
\includegraphics[width=0.35\linewidth]{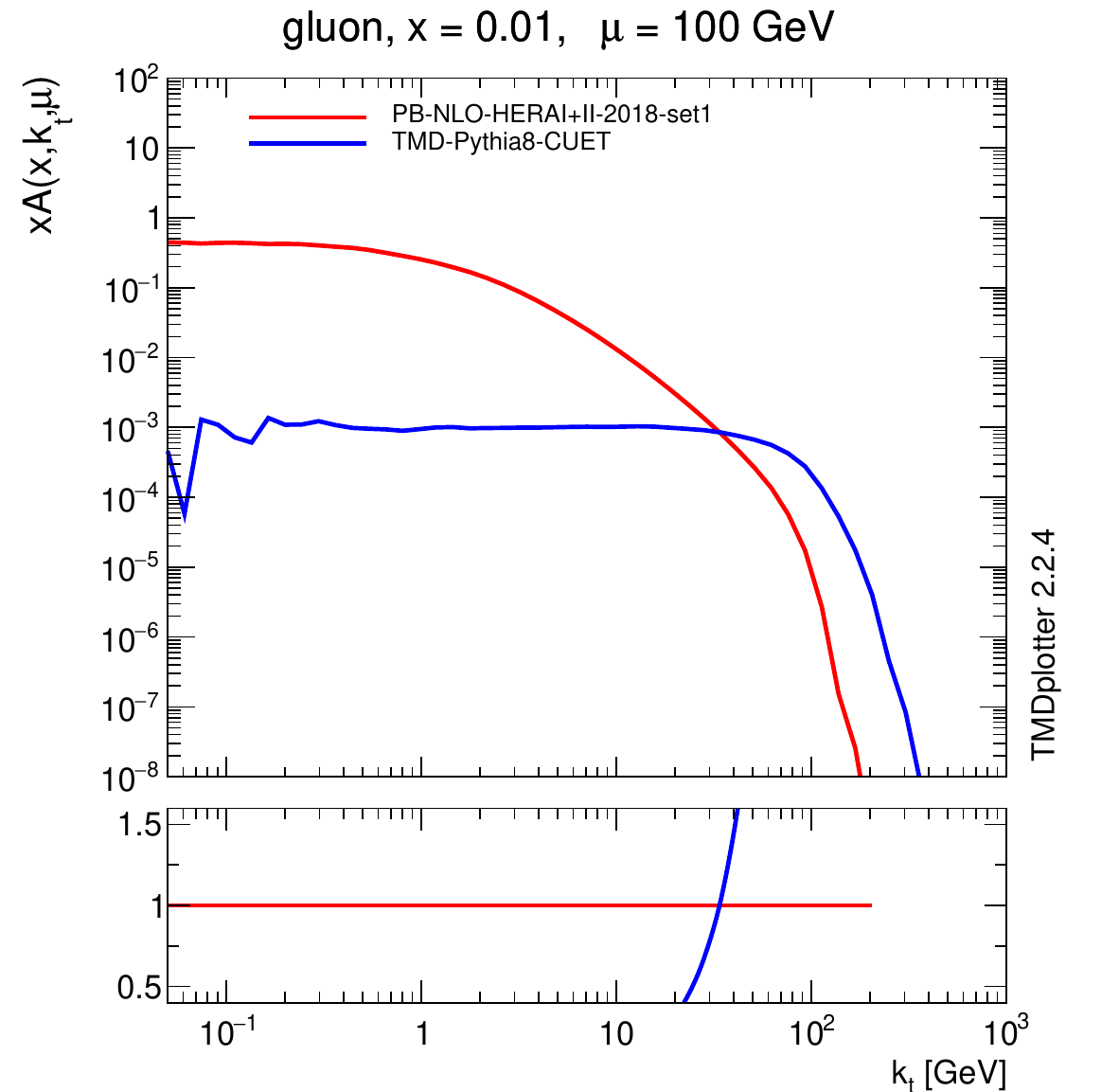}  
\includegraphics[width=0.35\linewidth]{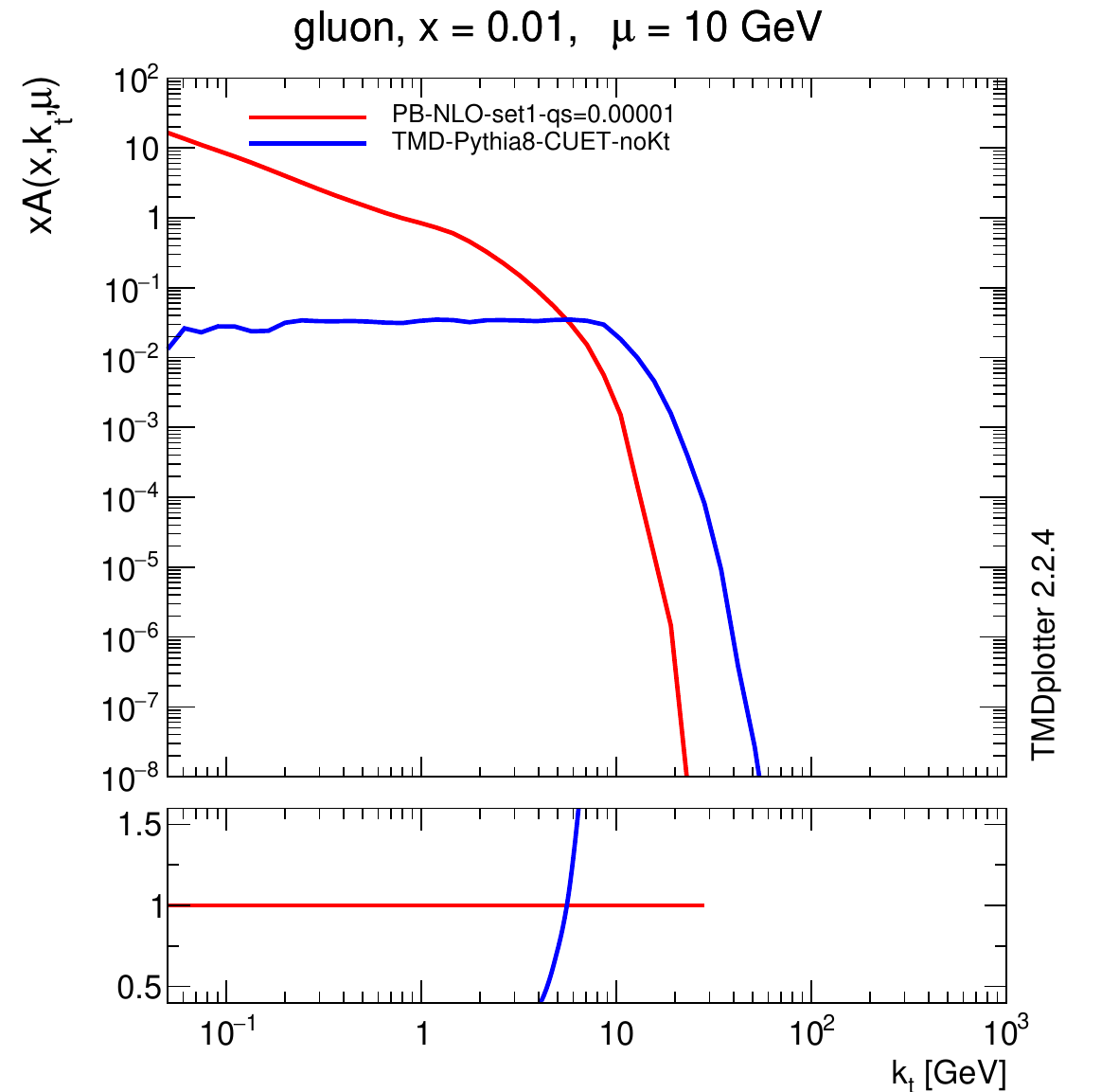} 
\includegraphics[width=0.35\linewidth]{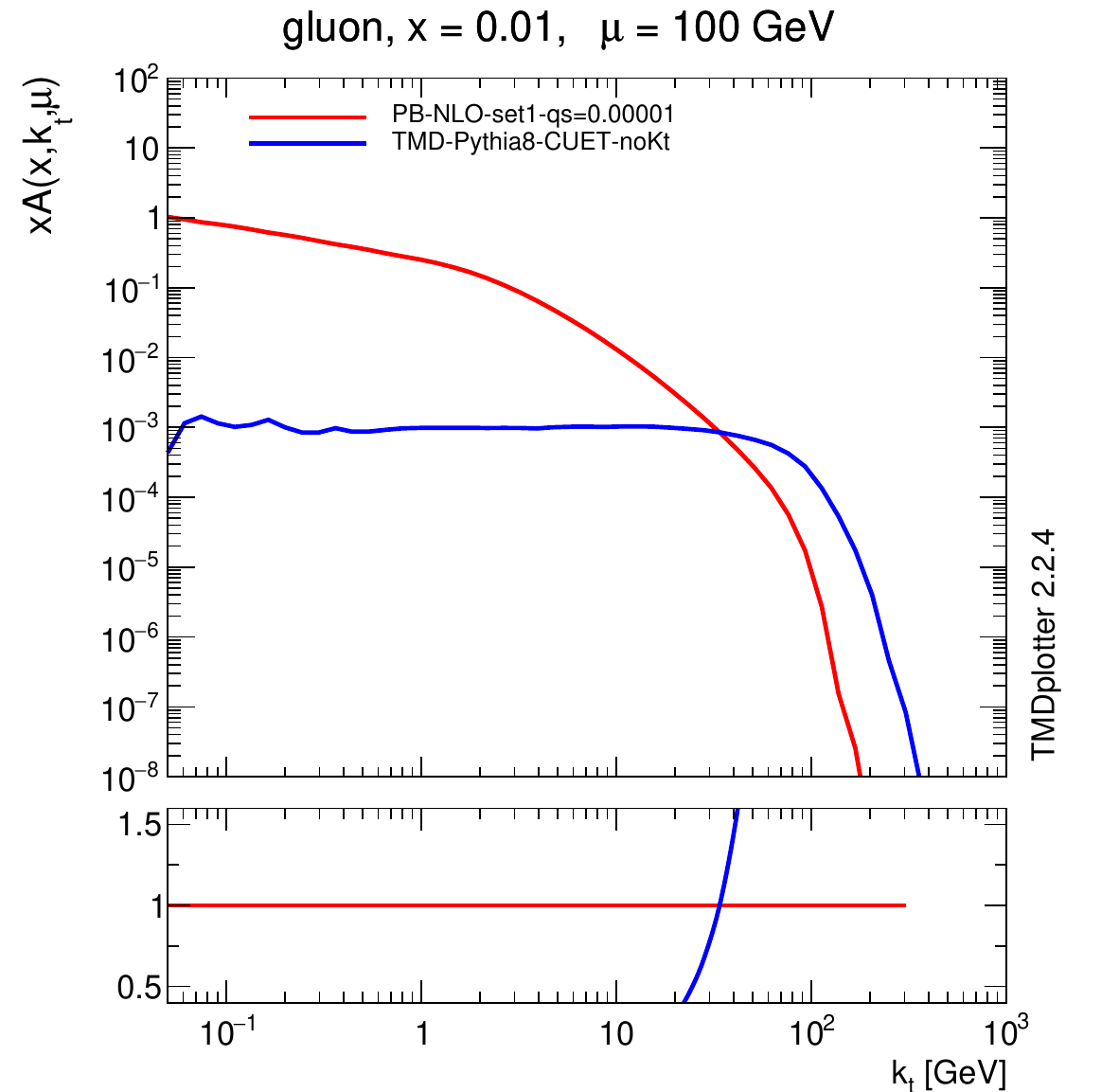}  
\caption {\small Transverse momentum distributions for gluons at two different scales $\mu=10 (100) $~\GeV\ (\PBNLOset~Set1), obtained from \protect\PBM -method~\protect\cite{Martinez:2018jxt} and \protect\pythia\ (CUET tune). 
 The upper row shows the distributions with the intrinsic-\kt\ distribution, and the lower row shows the distributions without it.}
\label{fig:PB-P8CUET}
\end{figure}

\begin{tolerant}{8000}
In Fig.~\ref{fig:PB-P8CUET} (upper row) we show a comparison of TMD distributions obtained from \PBNLOset~Set1 \cite{Martinez:2018jxt} with those from space-like parton showers of \pythia . 
The distributions differ significantly, as expected, due to the different ordering conditions used. For better comparison between the true parton shower and the \PBM -evolution,  we show the distributions without including any intrinsic-\kt\ contribution  in Fig.~\ref{fig:PB-P8CUET} (lower row).
\end{tolerant}

\section{The \pythiaPB\ method in \pythia}

The \pythiaPB\ method is developed using the \PBM\ method with
collinear and TMD parton densities, combined with the \pythia\ parton shower
machinery.  The basic ingredients of the \PBM\ collinear and TMD
parton densities are angular ordering and the choice of the scale in
\as. We reinterpret the evolution scale in \pythia to be
$p_{\perp\rm evol}=p_\perp/(1-z)$ rather than $p_\perp$
(for a technical description see Appendix~\ref{P8mod}) in a way such
that the starting scale, $\mu_0$, can be identified with
$p_{\perp\rm cut}$ (and setting $p_{\perp0}^2=0$, to
avoid the smooth suppression described in
section~\ref{sec:pythia-isr}). The quark masses are chosen according
to the \PBM\ distributions.
A complete list of the parameters used is given in Appendix~\ref{P8param}. 

In eq.~(\ref{Suda}), the integral over $z$ is limited by \zM . In
the DGLAP framework, $\zM = 1$, whereas in a numerical calculation $\zM \neq 1$ due to
the presence of $1/(1-z)$ poles in the splitting functions. In parton shower
approaches, it is often argued that $z$ is limited by kinematics, and
by requiring a minimum transverse momentum of the emitted parton one
obtains a limit on $\zM < 1$.
However, as argued in
Ref.~\cite{Mendizabal:2023mel} (and shown explicitly in
Ref.~\cite{Bubanja:2024puv} for the case of the \pt\ spectrum of DY
pairs) soft gluons with $z \to 1$ play an important role especially in
the small \kt -region. In \PBM\ Set1, $\zM = 0.99999$ is used, and the
same value is also applied in the \pythia\ studies presented here.

In order to comply with the treatment of heavy flavors in \PBM, which
follows the Variable Flavor Zero Mass (VFZM) scheme, any special heavy
flavor treatment in the \pythia\ parton shower has been
disabled. 

\subsection{Effective TMDs from \pythia -\pythiaPB\ with fixed \boldmath\as }

We begin by calculating \PBM\ collinear and TMD distributions with  \updfevolv ~\cite{Jung:2024uwc} at LO and NLO (keeping all other parameters as in \PBNLOset\ (Set2) but without intrinsic \kt -distribution). To specifically focus on the splitting functions, we apply a  fixed value of $\as = 0.13$.

\begin{figure}[t]
\centering
\includegraphics[width=0.35\linewidth]{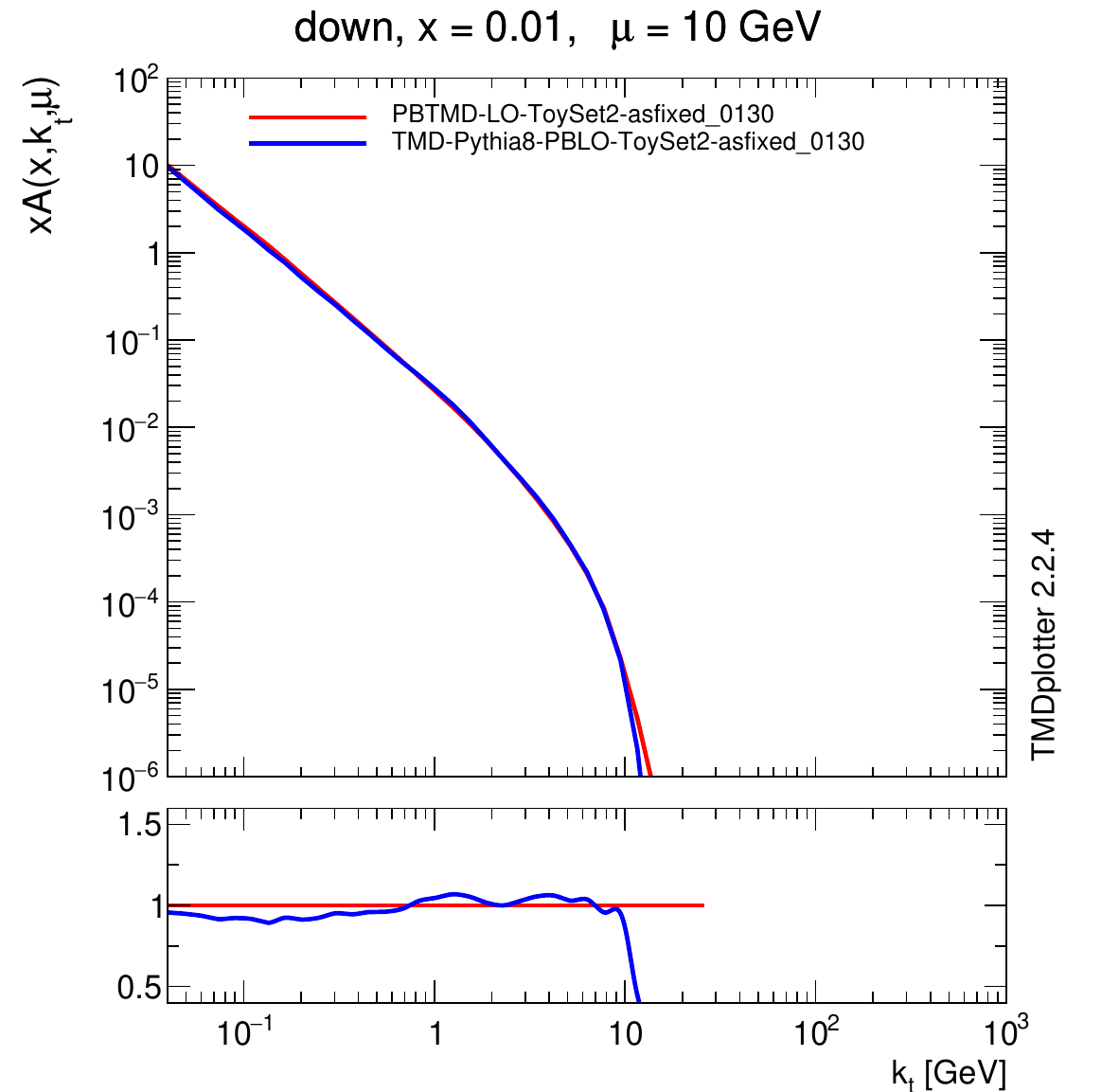} 
\includegraphics[width=0.35\linewidth]{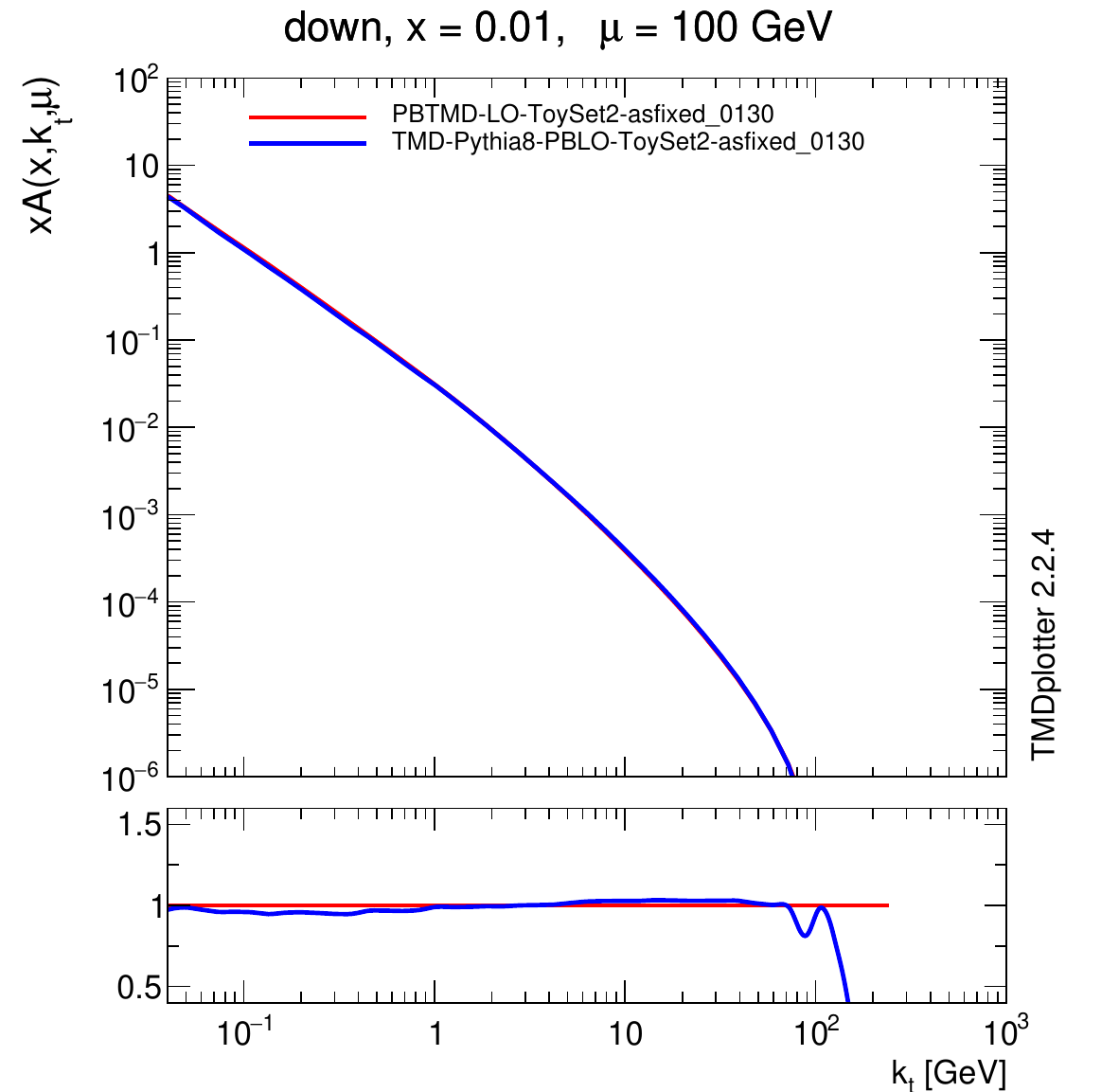}  
\includegraphics[width=0.35\linewidth]{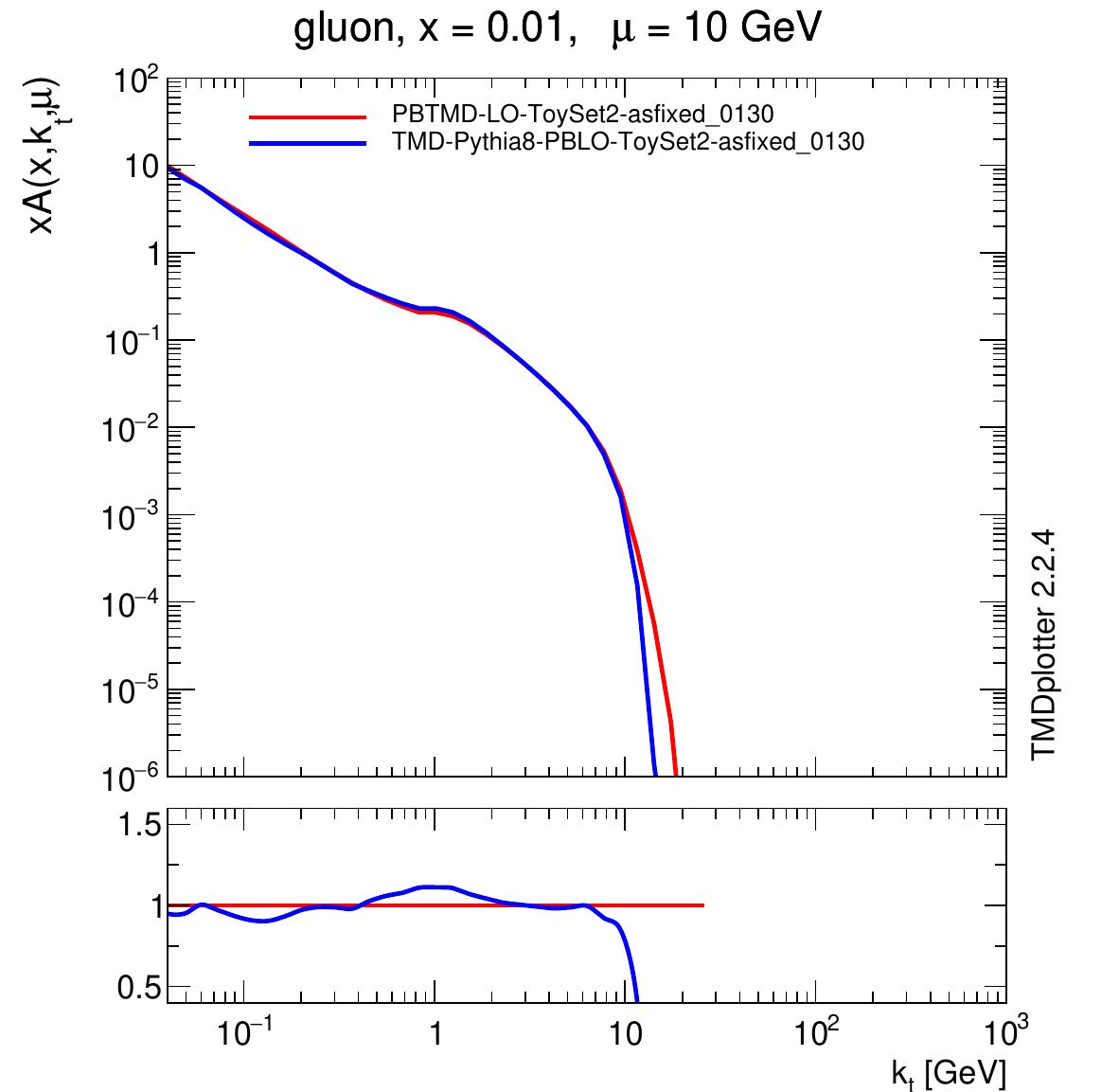} 
\includegraphics[width=0.35\linewidth]{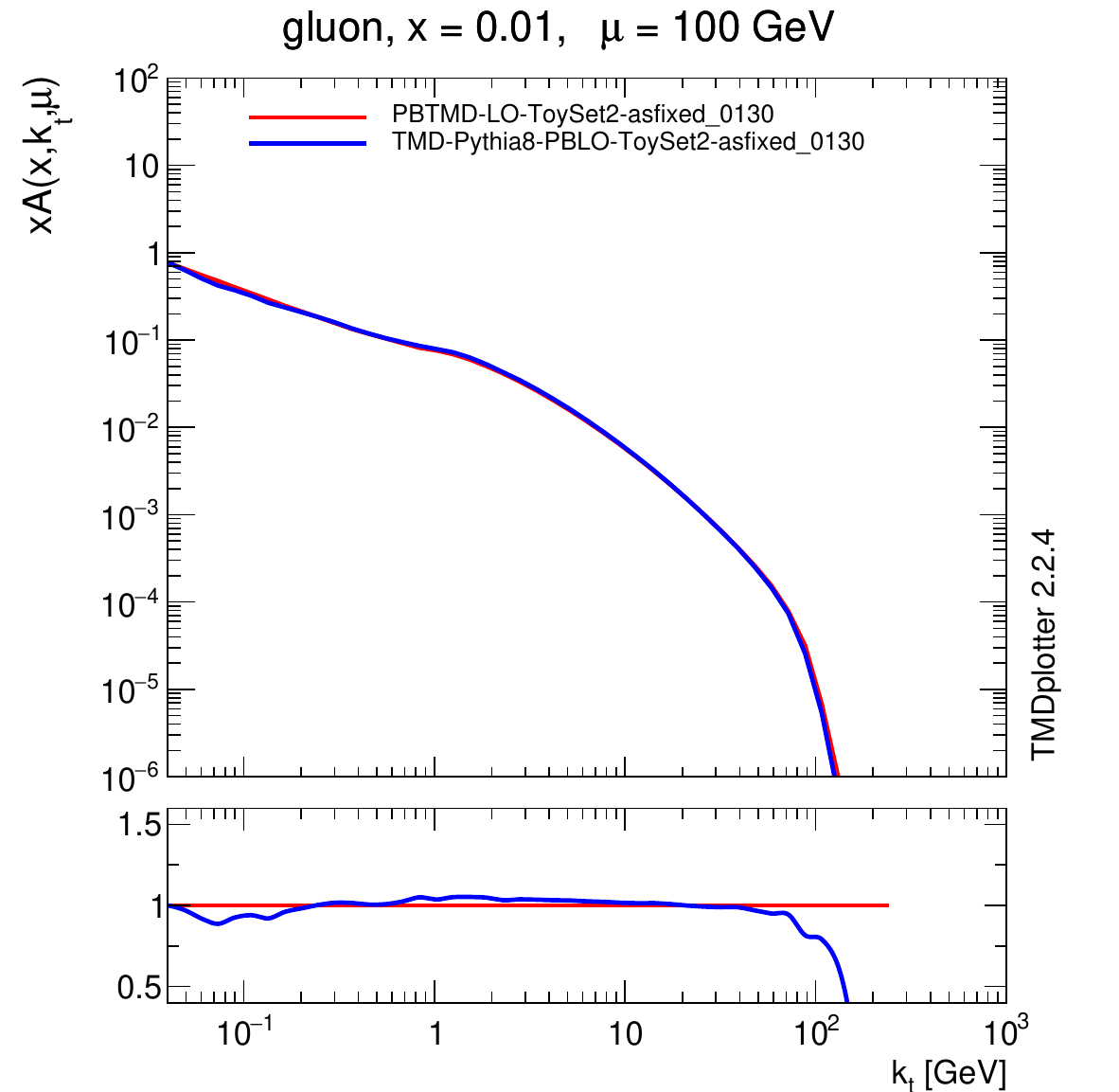}  
\cprotect\caption {\small Transverse momentum distributions for gluons and down-quarks at $\mu=10 (100) $\GeV , obtained from \PBToyset~Set2 evolved from a starting scale $\mu_0=1.18$ \GeV\  and \protect\pythia\ applying angular ordering and \verb+pTmin =1.18+ \GeV.
The predictions are obtained  at LO (with fixed $\as = 0.130$).
}
\label{fig:LOToy_fixed_as}
\end{figure}

In Fig.~\ref{fig:LOToy_fixed_as} we compare the calculations of  LO \PBM -TMD distributions with the one obtained from the \pythia\ parton shower applying \pythiaPB\  for down quarks and gluons at two different scales $\mu=10 (100)$ \GeV.  The results are in very good agreement, indicating that the \pythiaPB -method successfully reproduces distributions obtained from the \PBM -method.

\begin{figure}[t]
\centering
\vskip -1cm
\includegraphics[width=0.35\linewidth]{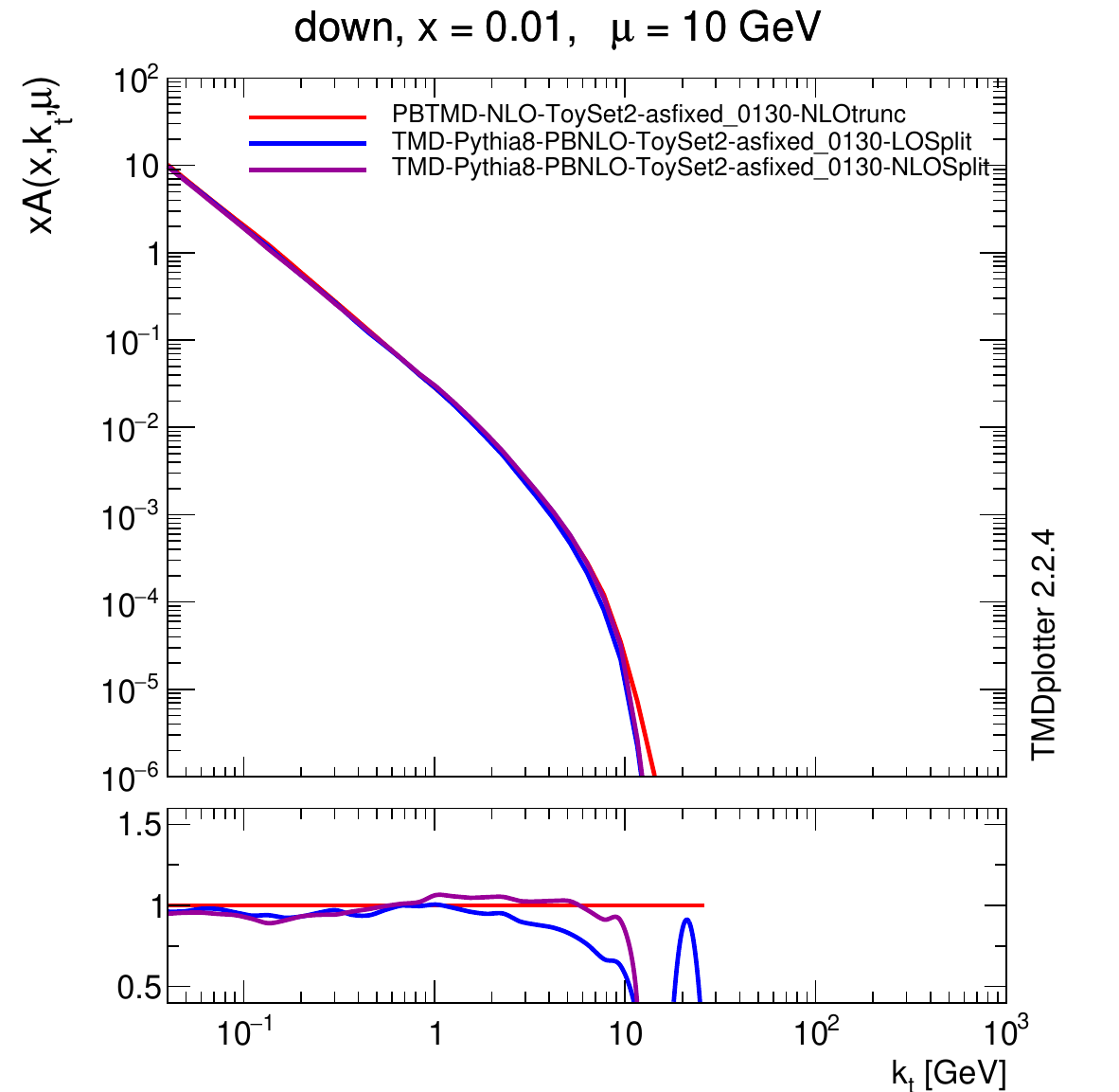} 
\includegraphics[width=0.35\linewidth]{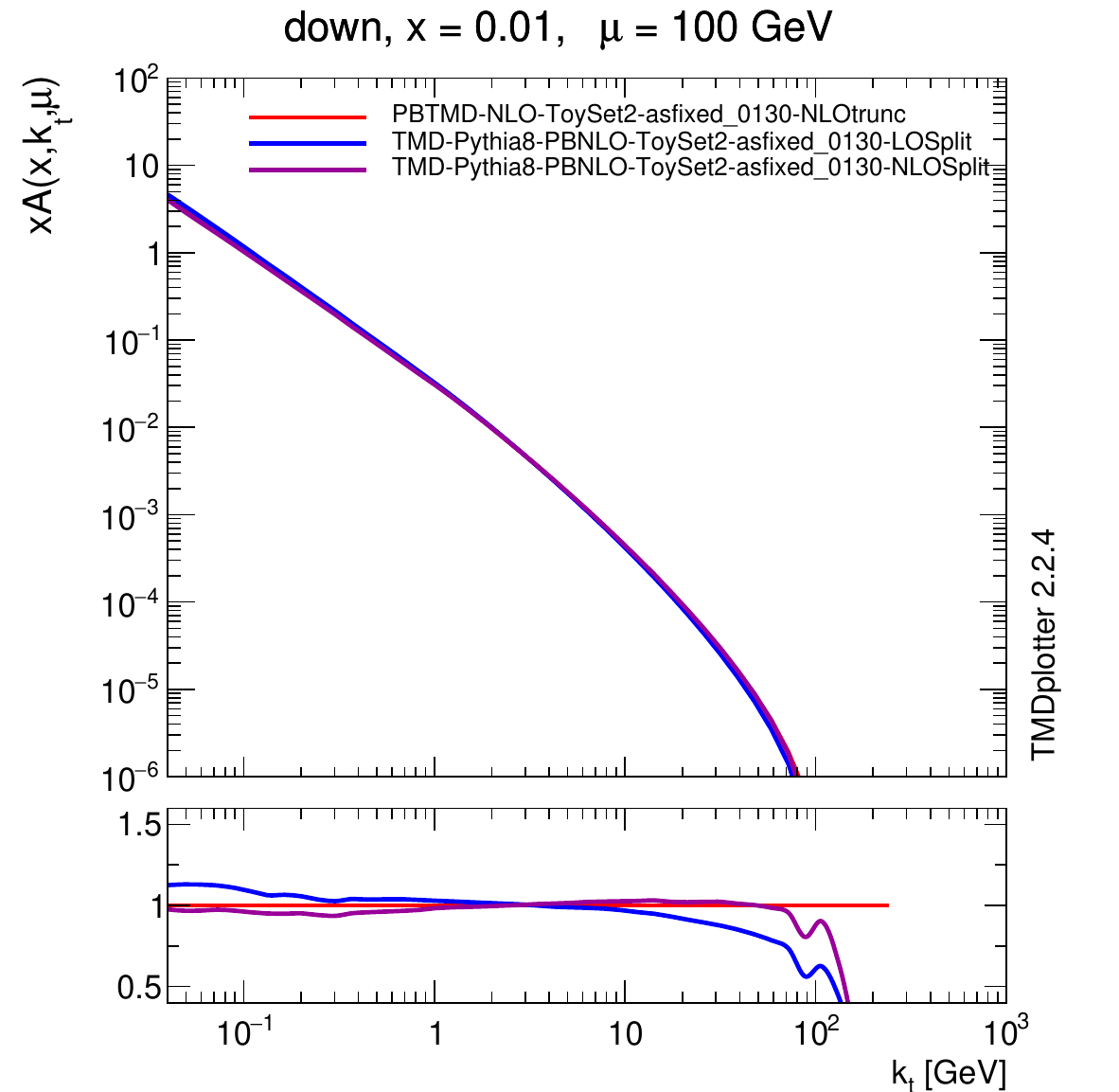}  
\includegraphics[width=0.35\linewidth]{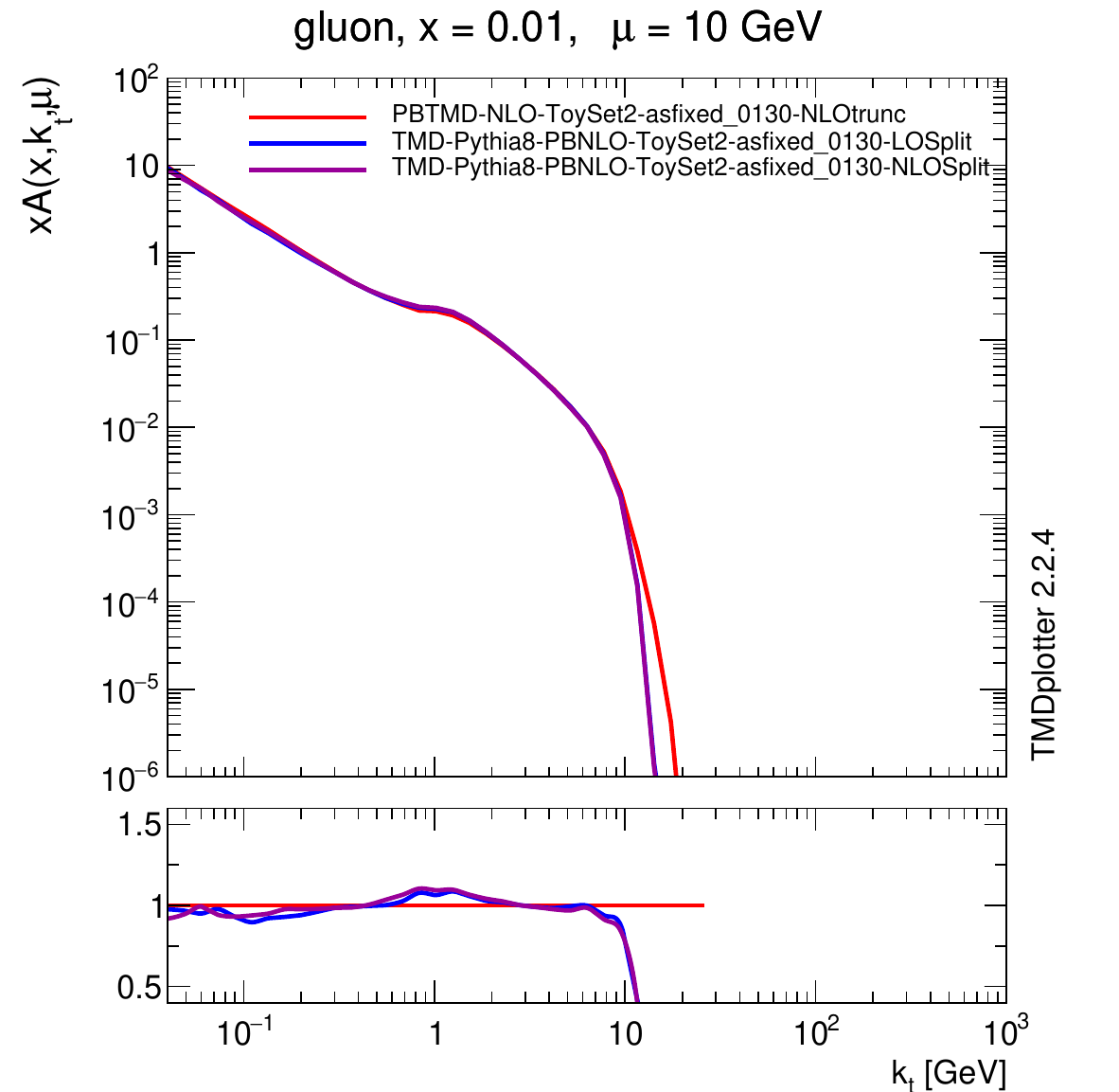} 
\includegraphics[width=0.35\linewidth]{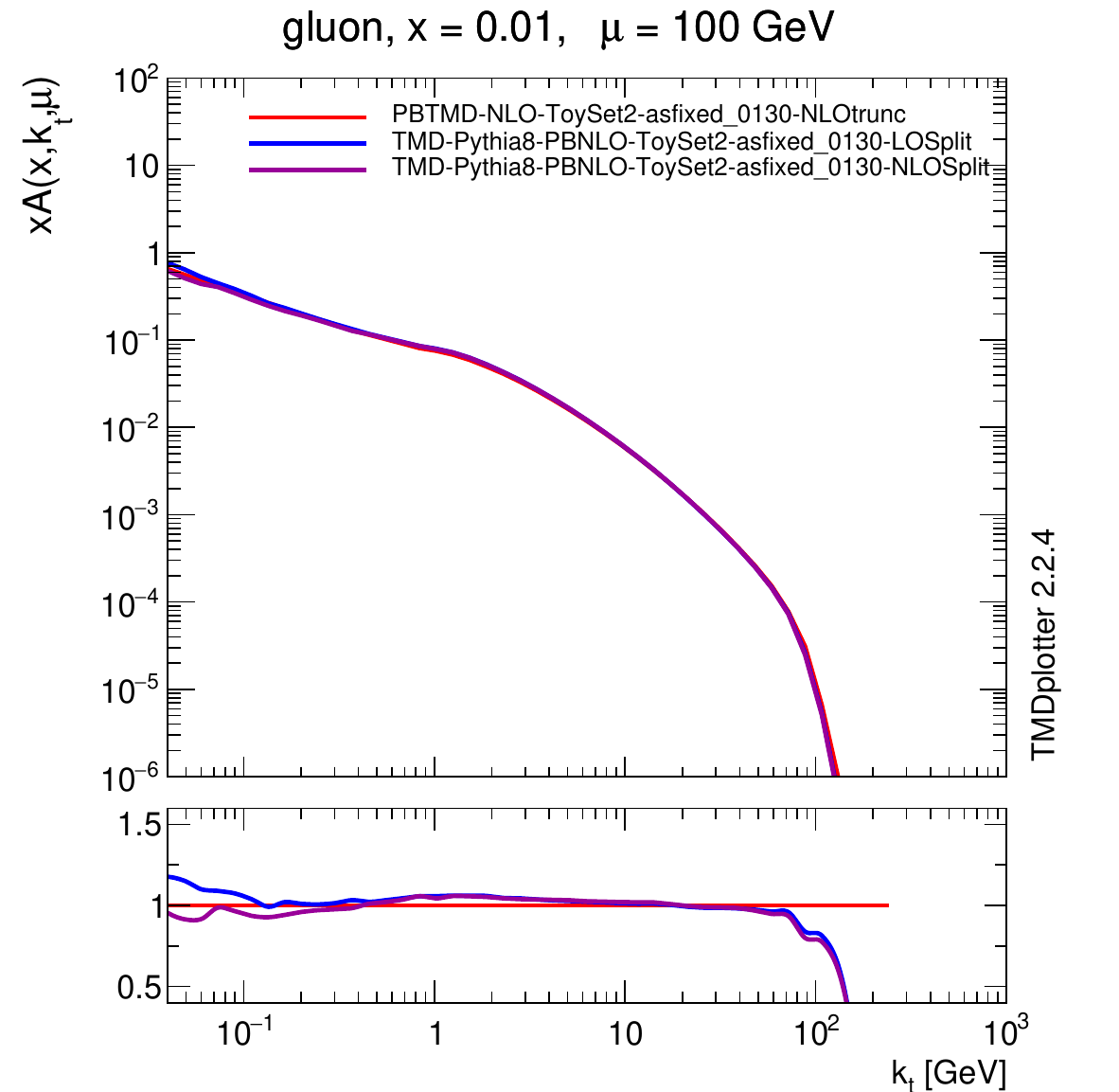}  
\cprotect\caption {\small Transverse momentum distributions for gluons and down-quarks as $\mu=10 (100) $\GeV. The red line is obtained from \protect\updfevolv\  (PBTMD-ToySet2) evolved from a starting scale $\mu_0=1.18$ \GeV\  using NLO splitting functions (for details see text).  The predictions using the \pythia-\pythiaPB with LO splitting functions are shown in blue, those with NLO splitting functions are shown in purple. In all cases fixed $\as = 0.130$ is applied.
}
\label{fig:NLOToy_fixed_as}
\end{figure}

We now investigate distributions obtained with NLO \PBM -TMD distributions. In Fig.~\ref{fig:NLOToy_fixed_as}, a comparison is 
presented between calculations using \updfevolv\ with NLO splitting functions (in toy mode with fixed \as ) and predictions from the \pythia -\pythiaPB\ parton shower with standard LO splitting functions (blue lines). Obviously, significant differences in the TMD distribution for quarks are observed, which also illustrates the inconsistency using NLO parton densities with LO splitting functions inside the parton shower.

We have implemented the full NLO splitting functions (taken from QCDnum~\cite{Botje:2010ay}) to be used in the  \pythia -\pythiaPB\ parton shower. In the ISR simulation only about $ 0.1 \%$ branchings come with negative weights, mainly coming from the region of large $z$ and low $\pt < 1$~\GeV.\footnote{At this stage we ignore negative parts of the splitting function. The correct treatment will be discussed later.}  The purple line in Fig.~\ref{fig:NLOToy_fixed_as} shows the predictions from the \pythia -\pythiaPB\ with NLO splitting functions, restricted to channels that also appear at LO (labeled as NLOtrunc).\footnote{We have explicitly checked and confirmed that the additional channels at NLO have a negligible effect on the TMD distributions. }  The agreement with the NLO \PBM -TMD distributions is significantly improved, highlighting the mismatch when different orders of splitting functions are used in the evolution and in the parton shower.

\section{\pythiaPB\ at NLO}

In the previous section we have shown with the toy model that the \pythia -\pythiaPB\ can reproduce the TMD distributions very well  both at LO and NLO provided the corresponding order of splitting functions is applied.  We can now discuss predictions obtained with the available \PBNLOset\ distributions. A comparison with the LO distributions is shown in Appendix~\ref{TMDLO}.
We use  \PBM -TMD distributions at NLO obtained in Ref.~\cite{Martinez:2018jxt} including an intrinsic-\kt\ distribution with width $q_s=0.5$~\GeV .

\subsection{\boldmath\as\ at  NLO}
\begin{figure}[t]
\centering
\vskip -0.2cm
\includegraphics[width=0.9\linewidth]{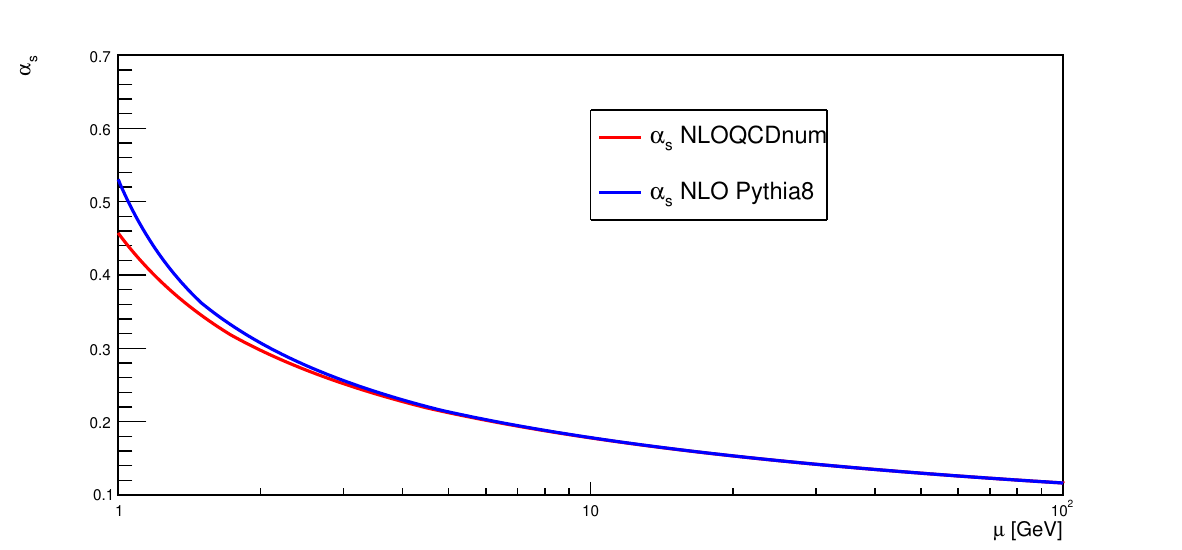} 
\cprotect\caption {\small The strong coupling as a function of the scale $\mu$  (with $\asmz = 0.118$) obtained from \pythia\ and \qcdnum .  }
\label{fig:alphas}
\end{figure}

The \PBNLOset\ distributions were obtained using \as\ and the splitting functions as implemented in \qcdnum . While \as\ at LO is trivial (provided the same value of $\Lambda_{qcd}$ and the same mass thresholds are used), differences in the evolution of \as\ at NLO show up, depending on which evolution scheme is used. Inside \pythia\ the evolution scheme from PDG~\cite{ParticleDataGroup:2006fqo} is applied, while \qcdnum\ uses a numerical integration instead of a parameterization. Differences between the two approaches are visible especially in the region of low scales, as shown in Fig.~\ref{fig:alphas}. Technically, the parametrization which is stored in the LHApdf file will be employed later.

\subsection{The treatment of negative contributions in NLO splitting functions}

A parton shower simulation is based on probabilities for radiation by
interpreting the Sudakov form factor eq.(\ref{Suda}) as a no-emission
probability. Technically, the generation is done by the so-called Veto
Algorithm, where the splitting functions can be conveniently
overestimated in a first step, thus underestimating the no-emission
probability, and then applying a veto in a secondary step to obtain
not only the correct emission probability but also the correct
no-emission probability (see  \cite{Sjostrand:2006za}). This
interpretation of the Sudakov form factor is, however, only valid for
LO splitting function. At NLO, the splitting functions are no longer
positive definite, and any interpretation in terms of probabilities is therefore 
excluded. We can overcome this problem by using LO splitting functions
in the generation, but introducing an extra accept--reject step in the
generation where an overall event weight is calculated from the ratio
of NLO/LO as described in Refs.~\cite{Mrenna:2016sih,Lonnblad:2012hz}. This ensures that the end result corresponds to using NLO both for the splitting
functions and the Sudakov form factors. This procedure is described in
more detail in Appendix~\ref{P8mod}, and will produce fluctuating, and
sometimes negative, event weights, which affect the statistical
significance.

\subsection{Full TMD distributions obtained with \pythiaPB\ at NLO}
We are now in a position to compare the complete NLO distributions
obtained from \pythia\ \pythiaPB\ with the TMD distributions of
\PBNLOset , which we used here as a test-case to show the consistency
of the whole procedure. The method itself is universal, and can be
also applied to any other collinear parton density at NLO.

Two things are worth noting here. The NLO splitting functions contain
$1\to3$ type splittings that are integrated over, but the modified
\pythia will emit only one parton, so only the (backward)
evolution that is corrected to NLO, while the partons radiated into
the final-state are still described correctly only at LO. This also
means that it is straightforward also to go to NNLO evolution, keeping
in mind that the final-state emissions are still at LO.

\begin{figure}[t]
\centering
\vskip -0.2cm
\includegraphics[width=0.35\linewidth]{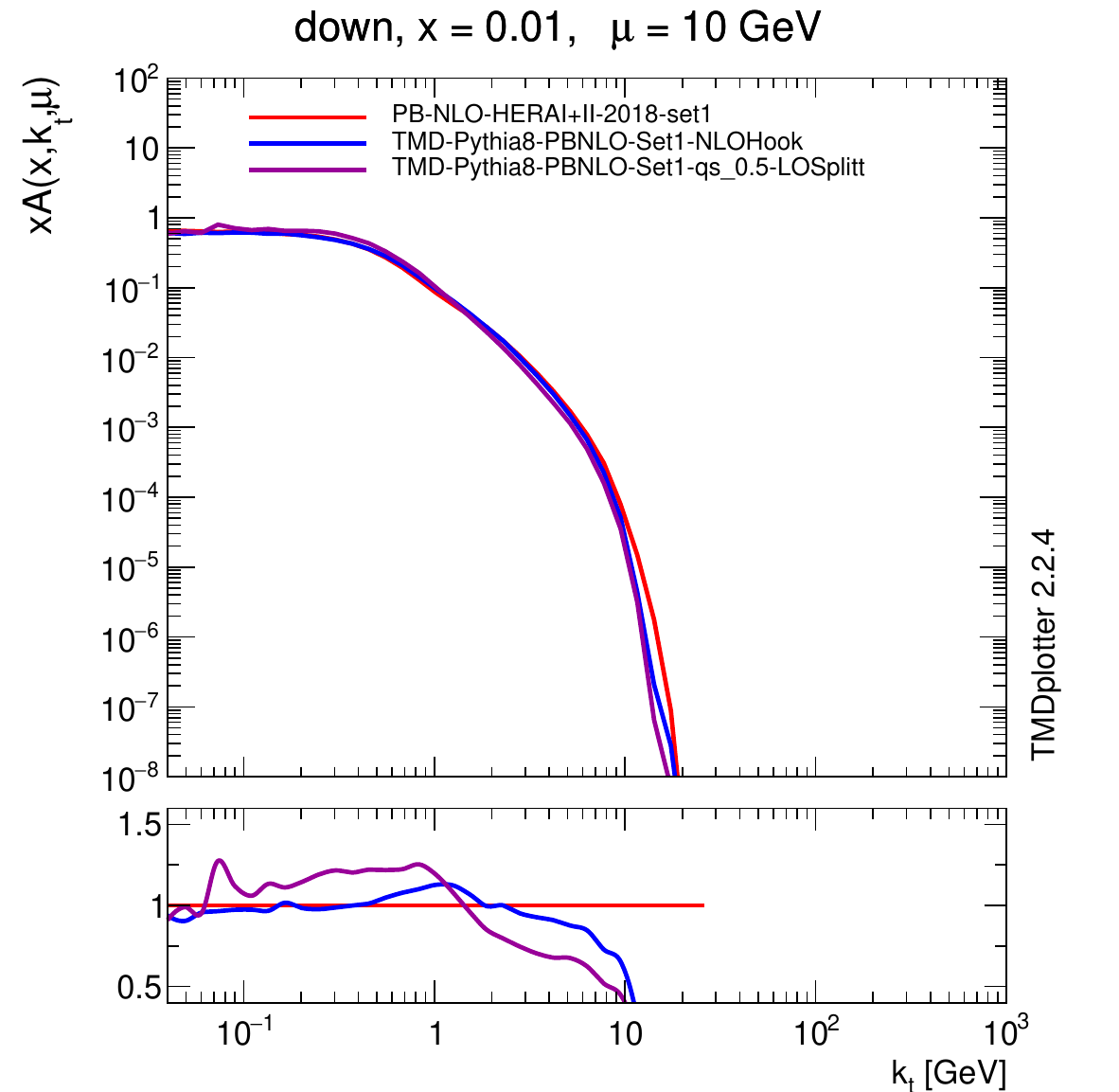} 
\includegraphics[width=0.35\linewidth]{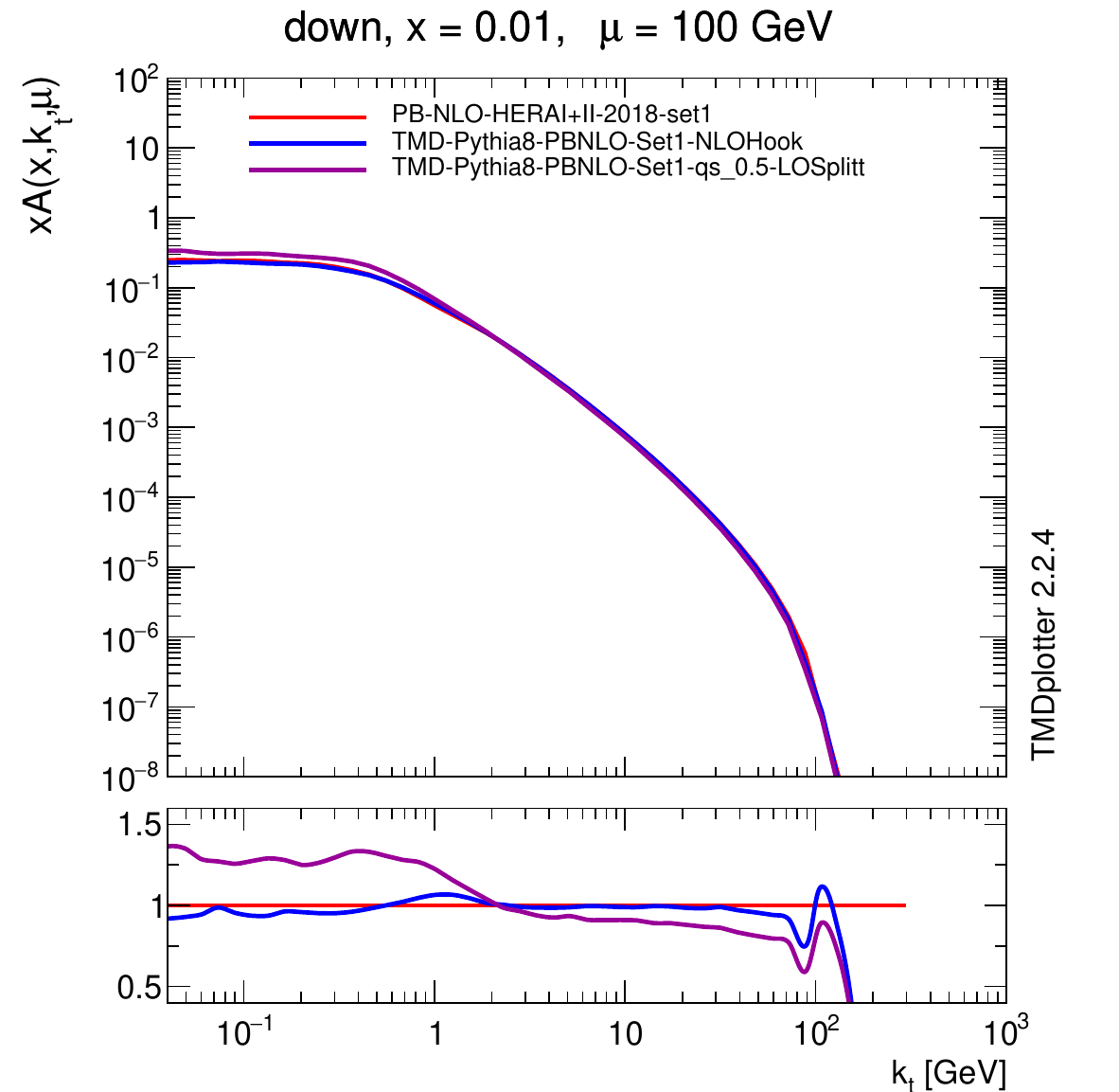}  
\includegraphics[width=0.35\linewidth]{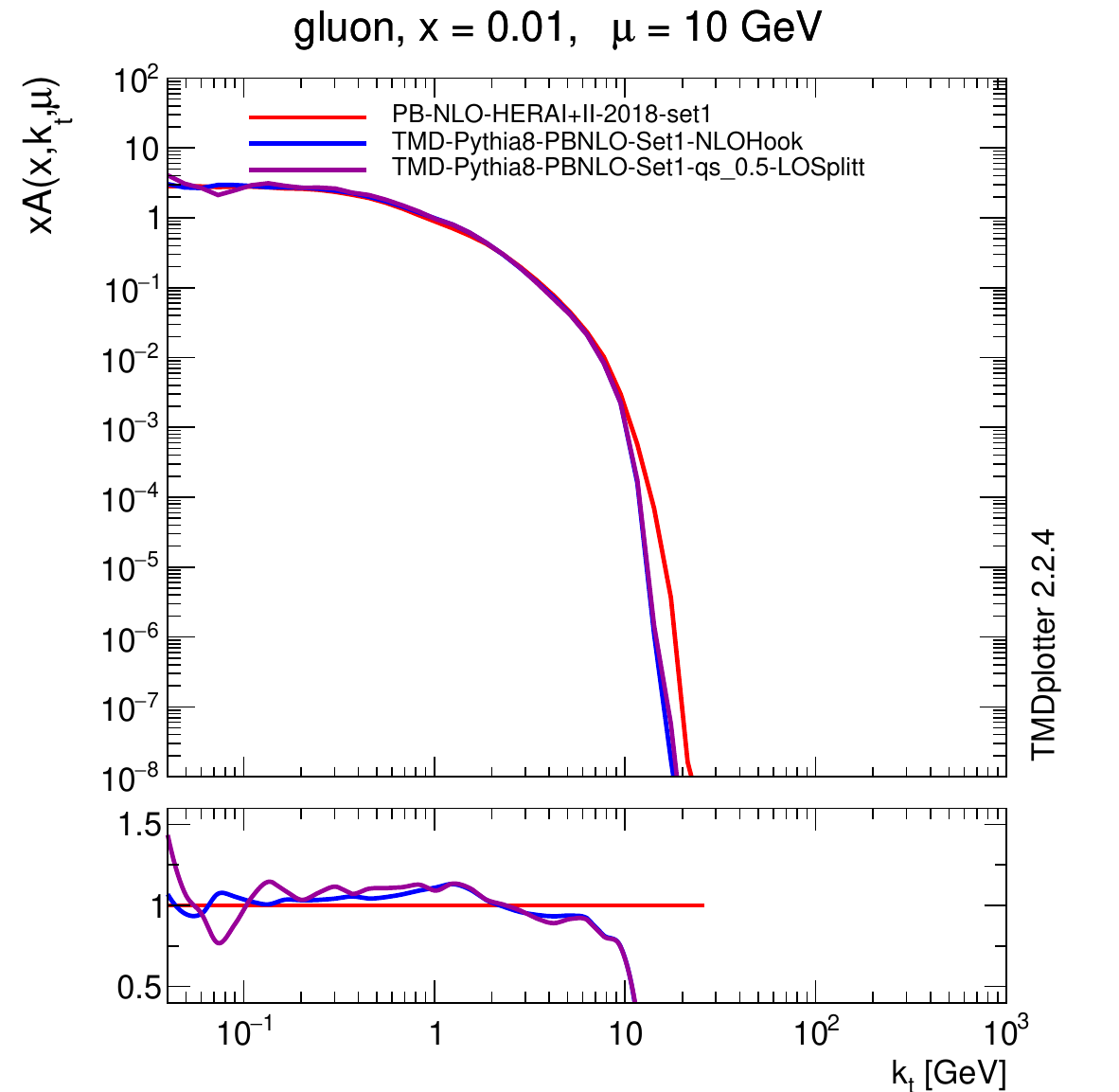} 
\includegraphics[width=0.35\linewidth]{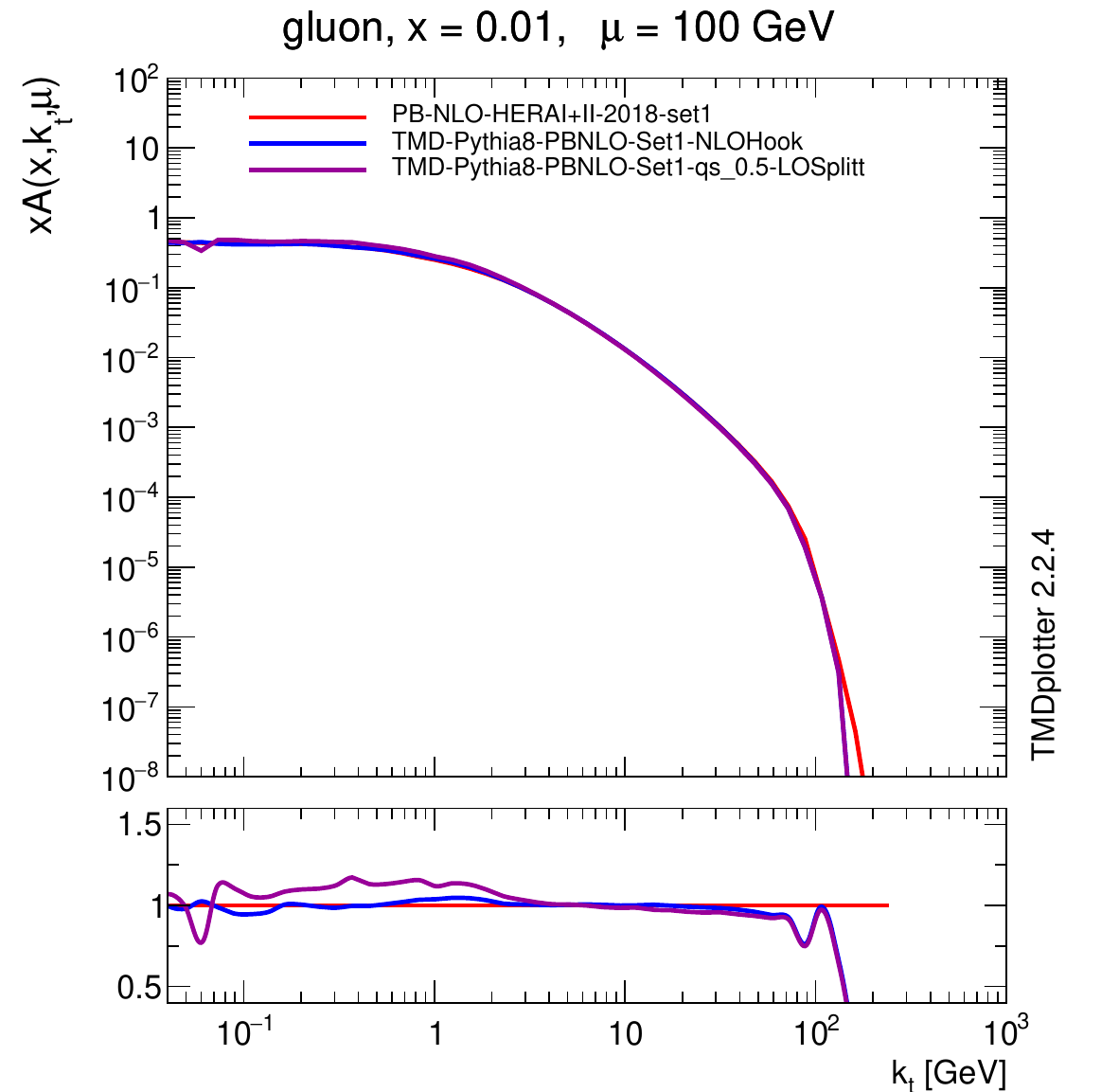}  
\cprotect\caption {\small Transverse momentum distributions for gluons and down-quarks as $\mu=10 (100) $\GeV , obtained from \PBNLOset~Set1 evolved from a starting scale $\mu_0=1.38$ \GeV\ and \protect\pythia\ \pythiaPB\ applying  \verb+pTmin=1.38+ \GeV .
The \PBNLOset\ predictions are obtained at NLO (with NLO $\asmz = 0.118$ obtained from the parmeterization in the LHApdf set).  The blue line is obtained from  \pythia\  \pythiaPB\ with NLO splitting functions. The purple curve shows the prediction using LO splitting functions but still with NLO \as .
}
\label{fig:PB-P8Set1-NLO-noKT-AngOrd}
\end{figure}

\paragraph{ \PBNLOset~Set1 conditions}
In  \PBNLOset~Set1 angular ordering is applied, but the scale in \as\ is set to the evolution scale (as done in all DGLAP-based collinear parton densities).  
In Fig.~\ref{fig:PB-P8Set1-NLO-noKT-AngOrd}, we  compare \PBNLOset~Set1 distributions, calculated with NLO splitting functions and NLO \as\ (as in Ref.~\cite{Martinez:2018jxt}), to distributions obtained from the parton shower in \pythia\ \pythiaPB , applying angular ordering and the \PBM~Set1 conditions. 
Here, we set \verb+pTmin=1.38+~\GeV\ , corresponding to the starting scale of $\mu_0=1.38$~\GeV .
For comparison, we also show predictions of \pythiaPB\ when only LO splitting functions are used for ISR. 
The distributions for the down quark and gluon are presented at different scales of $\mu=10~ (100) $ \GeV . The distributions agree very well if NLO splitting functions and  consistent \as\ values are used. Significant differences are observed, when only LO splitting functions are applied in the initial-state shower.

\paragraph{ \PBNLOset~Set2}
In \PBset~Set2, in addition to angular ordering, the scale in \as\ is set as the transverse momentum of the emitted parton, defined with $\qt = (1-z) q'$. At large $z$, \qt\ can become very small, requiring special treatment for \as\ at low scales; as in \PBM\ \as\ is frozen at $q_{cut} = 1$ \GeV . Details on how this is implemented in \pythia\ \pythiaPB\ are provided in Appendix~\ref{P8mod}.

Fig.~\ref{fig:PB-P8Set2-NLO-noKT-AngOrd} shows distributions for down quark and gluon at various scales $\mu$, applying \PBNLOset~Set2 at NLO. The blue curve shows the prediction using NLO splitting functions in \pythia\ \pythiaPB\ together with the consistent \as . The purple curve shows the prediction using the NLO calculation of \as\ as calculated in \pythia\, which is different at small scales from the one used in \PBNLOset , as shown in Fig.~\ref{fig:alphas}. It is interesting to observe that a consistent treatment of \as\ is required for a good description of the low \kt -part of the spectrum, especially for quarks.
The agreement of the simulation  \pythia\ \pythiaPB\ with the calculation of \PBNLOset~Set2 at NLO for the quark channel is remarkable.
The difference in the gluon channel arises from the use of different frame definitions when generating transverse momenta, as discussed in detail in Appendix~\ref{P8evol}, and can be treated as a systematic uncertainty related to the frame definition.

\begin{figure}[t]
\centering
\vskip -0.2cm
\includegraphics[width=0.35\linewidth]{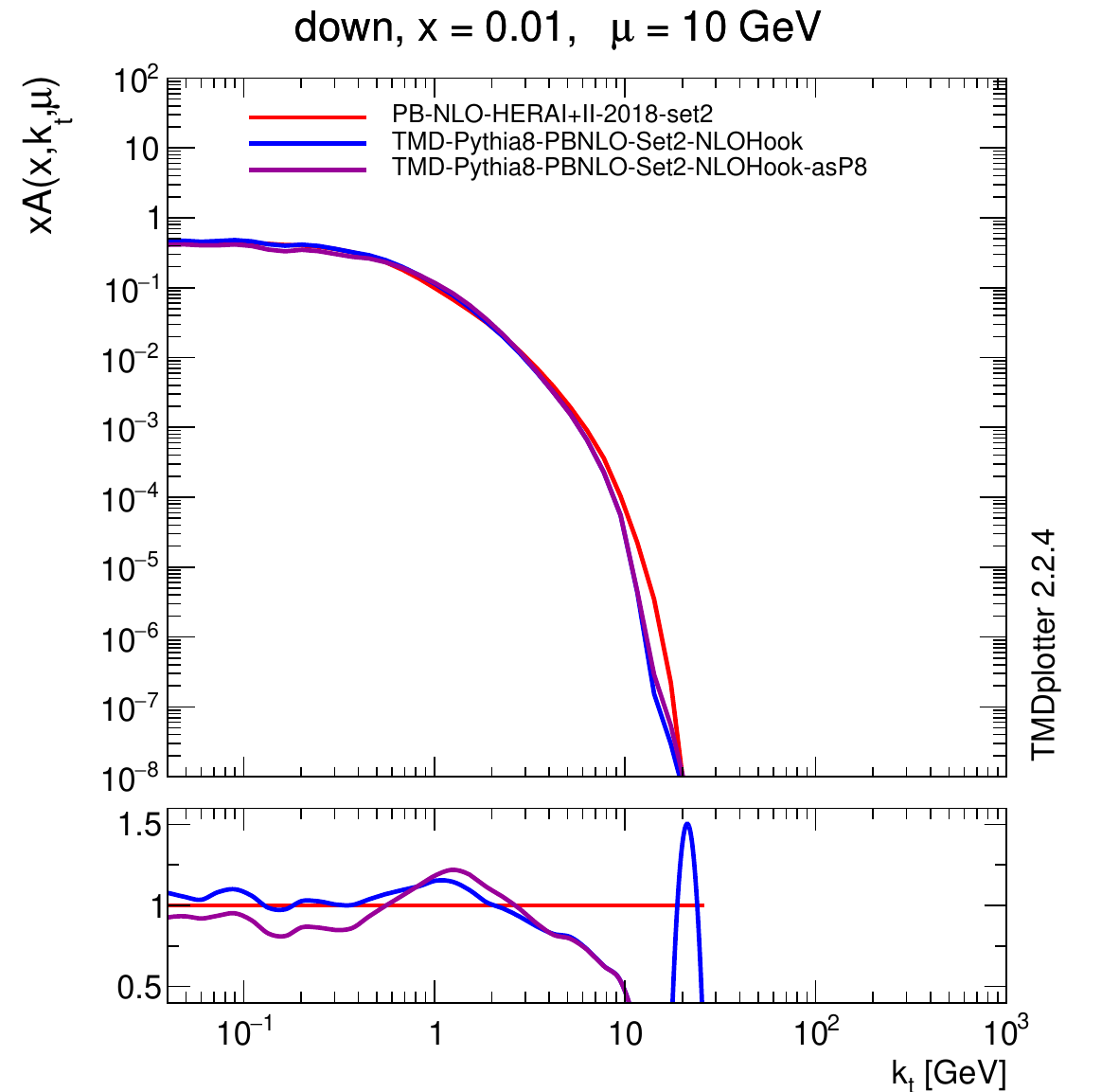} 
\includegraphics[width=0.35\linewidth]{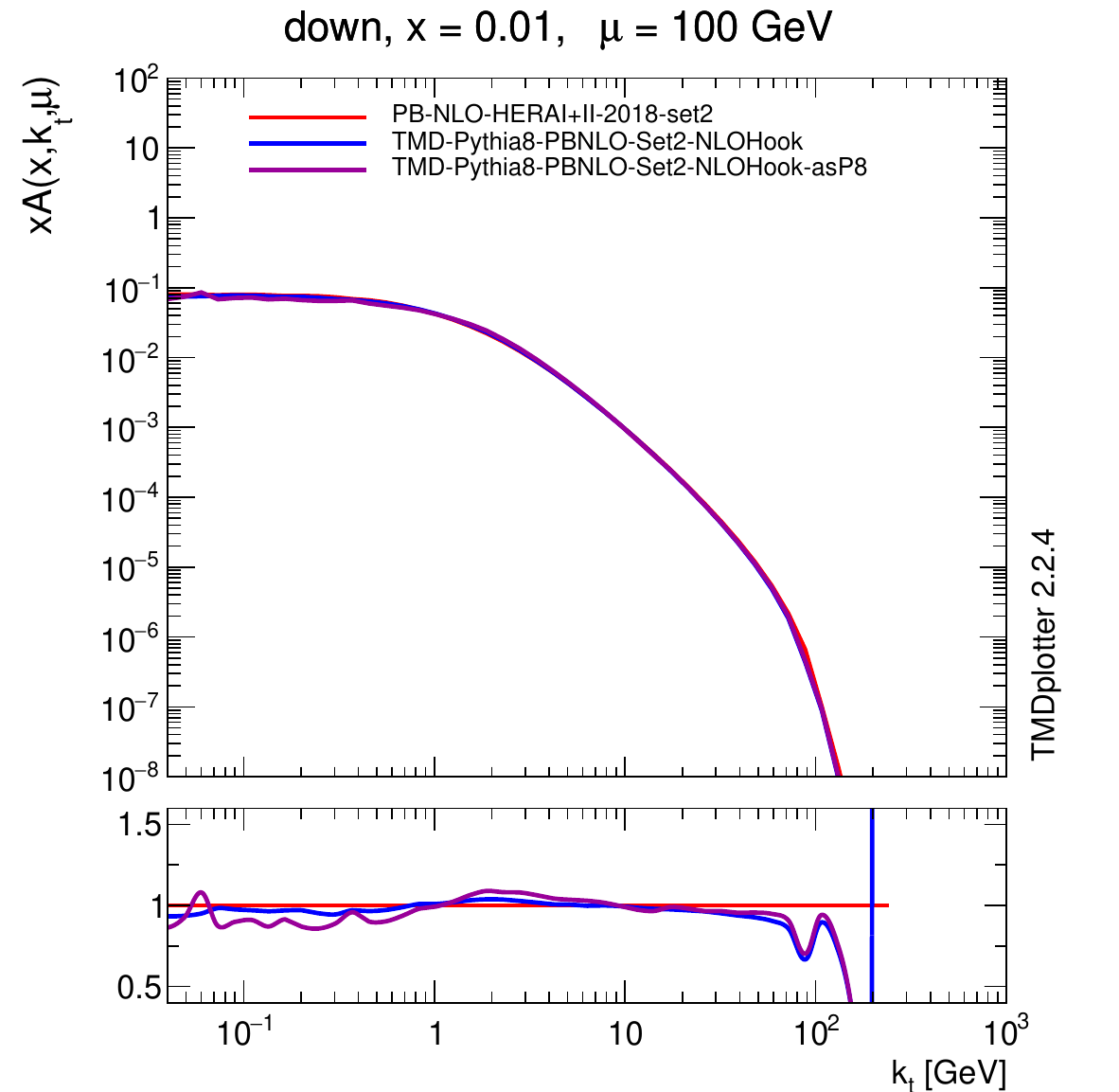}  
\includegraphics[width=0.35\linewidth]{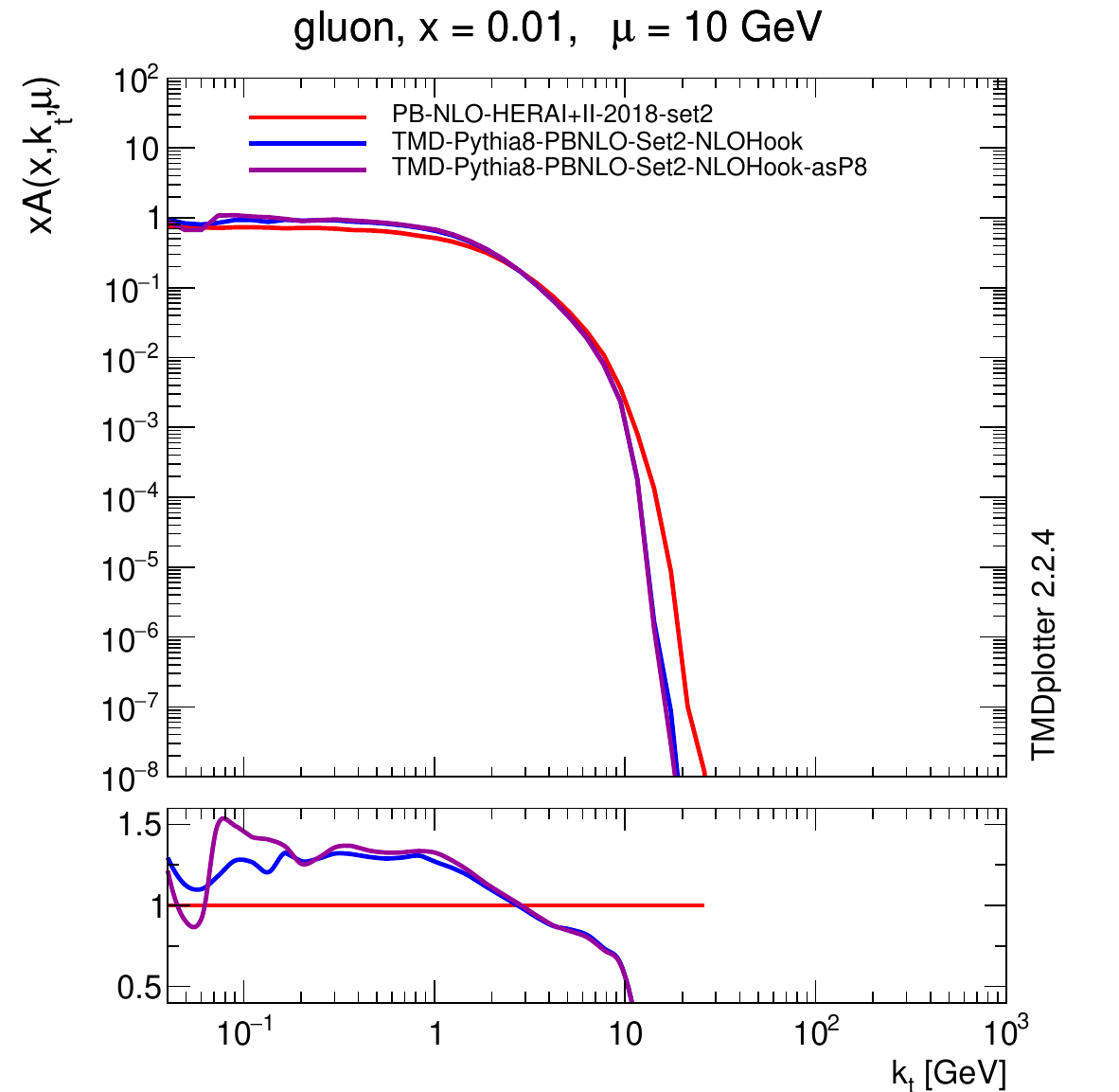} 
\includegraphics[width=0.35\linewidth]{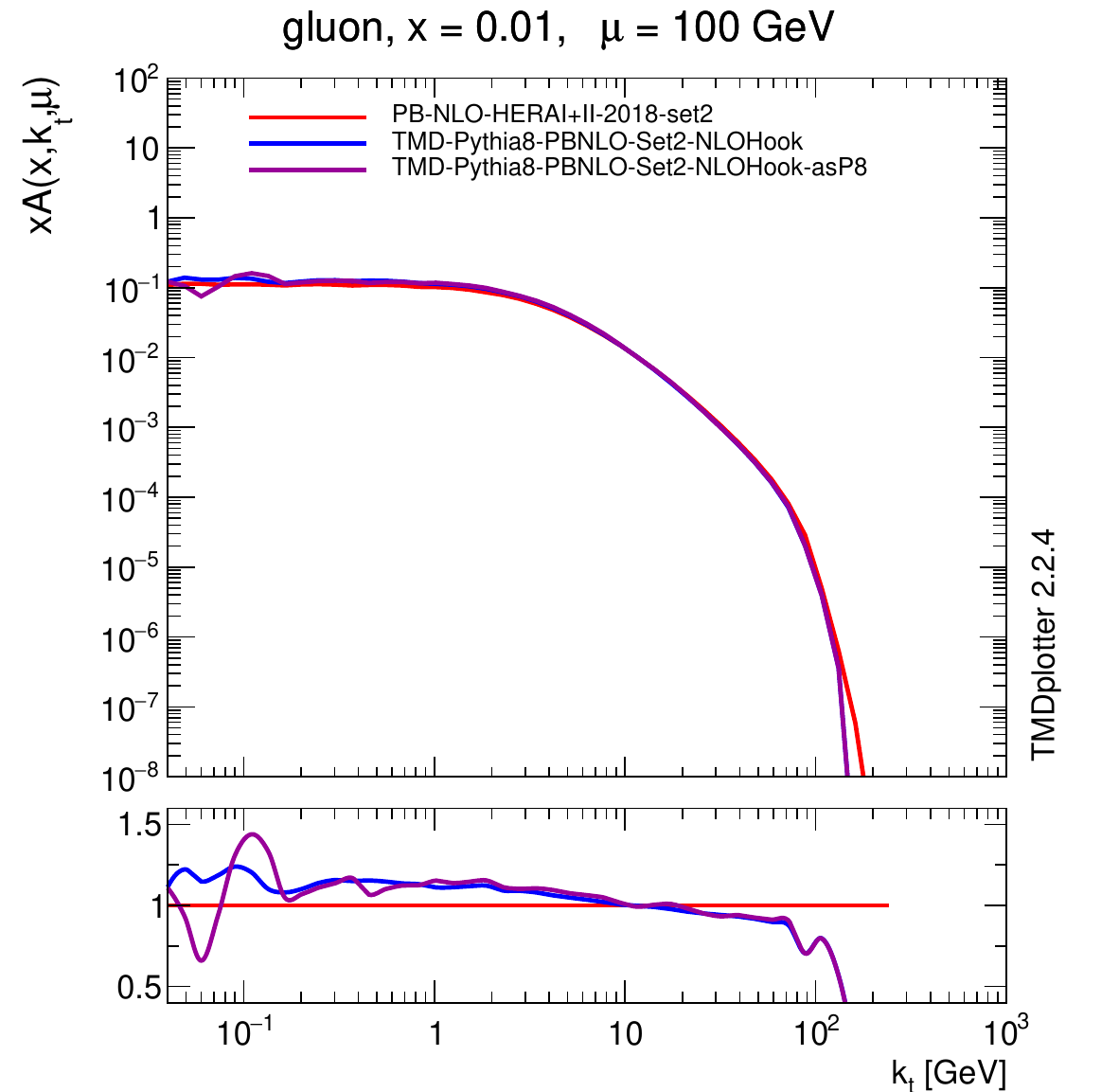}  
\cprotect\caption {\small Transverse momentum distributions for gluons and down-quarks as $\mu=10 (100) $\GeV , obtained from \PBNLOset~Set2 evolved from a starting scale $\mu_0=1.18$ \GeV\ and \protect\pythia\ \pythiaPB\ applying  \verb+pTmin=1.18+ \GeV .
The \PBNLOset\ predictions are obtained at NLO (with NLO $\asmz = 0.118$).  The blue line is obtained from  \pythia\  \pythiaPB\ with NLO splitting functions, the purple line shows the simulation when \as\ as calculated rom \pythia\ is used, which is different at small scales from the one applied in  \PBNLOset .
}
\label{fig:PB-P8Set2-NLO-noKT-AngOrd}
\end{figure}

\clearpage
\section {\label{sec4} Conclusions}

The main result of our study is that it is possible to construct an initial-state parton shower that is fully consistent with LO and NLO collinear parton densities.

In order to perform these studies, we have developed a method that allows us to construct Transverse Momentum Dependent parton densities from any parton shower event generator, a method we label as \pstmd . The Parton Branching (\PBM ) collinear and TMD parton densities were considered, since the TMDs provide a unique approach to study the parton branching processes in detail and in particular the effects from initial-state parton showers.

We found that using LO parton densities and LO splitting functions within the parton shower leads to consistent results: the TMD distributions obtained by the \PBM -approach in a forward evolution are identical to those obtained from the backward evolution parton shower with \pythia\ \pythiaPB , provided the same conditions are applied: angular ordering, kinematic limits, and the scale choice in \as .
This is already a big step forward in the understanding of parton showers and its relation to collinear parton densities.

A real breakthrough is acheived when considering TMD parton densities at NLO obtained with the \PBM -method. We could show that using NLO collinear parton densities but LO splitting functions within the parton shower leads to significant inconsistencies. Only by using the same evolution for \as\ together with NLO splitting functions can  consistent results be achieved. In order to achieve this, we also had to deal with negative  contributions from the splitting functions at large $z$ (and small \kt ). Applying a dedicated method of re-weighting an oversampled parton shower allowed us to treat these effects correctly. We also showed that the definition of the frame of reference in which  \kt\ is calculated matters; the frame is different in \pythia\ \pythiaPB\ and \updfevolv .

The   \pythiaPB\  method is universal and can be applied to any collinear parton density to obtain a consistent initial-state parton shower. The method is applicable at LO and NLO, as shown in this study, and can also be easily extended to NNLO.

\vskip 0.5 cm 
\begin{tolerant}{8000}
\noindent 
{\bf Acknowledgments.} 
S. Taheri Monfared acknowledges the support of the German Research Foundation (DFG) under grant number 467467041.
\end{tolerant}

\appendix

\section*{Appendices}

\section{Modifying ordering in and splitting functions \pythia \label{P8mod}}

The \pythiaPB\ code will be implemented as a plug-in to
\pythia, and can, in the meanwhile, be obtained upon request from the authors. The changes we have made to the \pythia code are not
substantial and are listed here for completeness.

\subsection{The ordering}
\label{sec:ordering}

The backward evolution in the default initial-state parton shower,
\texttt{SimpleSpaceShower}, in \pythia, is ordered in transverse
momentum, formally defined as (Ref.~\cite{Bierlich:2022pfr}[p.71])
\begin{equation}
 q^{\prime\, 2} = p_{\perp\rm evol}^2 =  (1- z) Q^2 - \frac{Q^4}{m^2_{ar}} \approx (1- z) Q^2 
\end{equation}
where $m^2_{ar}$ is the squared \textit{dipole mass} of the two
incoming partons on each side, which in our case is simply $m_B^2$ in
the first emission. The kinematics of the initial-state emissions are
calculated from the $z$ and $Q^2$ where $Q^2=-(p_b-p_c)^2$ (using the
notation in Fig.~\ref{fig:kine}), so it is straightforward to
reinterpret the evolution scale to be that of the angular scale
$p_\perp/(1-z)$ in a few places\footnote{Inside the code,
  $p^2_{\rm \perp evol}$ is given by the variable \texttt{pT2}. } and to
modify the relationship between $Q^2$ and $p^2_{\rm \perp evol}$ in the code,
\begin{equation}
  Q^2 \approx p_{\perp\rm evol}^2/(1-z) \qquad\Rightarrow\qquad
  Q^2 = (1-z)p_{\perp\rm evol}^2.
\end{equation}
The meaning of some parameters
will change, e.g., the soft suppression and hard cutoff discussed in
section~\ref{sec:pythia-isr} will now refer to the angular variable
rather than to the transverse momentum. So when we mention that we
have set, e.g., \texttt{pTmin = 1.18}, it corresponds exactly to
setting $\mu_0=1.18$~GeV in the TMD evolution.

\subsection{NLO splitting functions and \boldmath$\alpha_S$}
\label{sec:nlo-splitt-funct}

To modify the splitting functions and the $\alpha_S$ to conform to the
NLO functions in \qcdnum\ we use the event reweighting
technique in Ref.~\cite{Mrenna:2016sih}, implemented in a so-called
\texttt{UserHooks} plug-in class to \pythia. The plugin is accessed by
\pythia after each initial-state emission, inside the Veto algorithm,
and is asked whether the emission should be vetoed.

In our case we have artificially increased the fixed $\alpha_{S0}$
used in \pythia by a factor 2, and in the plug-in we then veto all
emissions with a probability 0.5. This will give the same results as
running without the plug-in but with an increased $\alpha_{S0}$, except for an
increase in running time. The trick is that we can now calculate an
event weight, $w_{\rm ev}$, that is used when filling histograms in
Rivet, to reweight our LO results to give the desired NLO
behaviour according to the following:
\begin{itemize}
\item For each suggested initial-state emission we calculate the
  NLO/LO ratios for the splitting functions and $\alpha_S$:
  \begin{equation}
    \label{eq:NLO-to-LO-ratios}
    r=\frac{P_{\rm NLO}(z)}{P_{\rm LO}(z)}\frac{\alpha_S^{\rm NLO}(k_\perp^2)}{\alpha_{S0}}
  \end{equation}
\item With probability 0.5 we will accept the emission, and update the event weight
  \begin{equation}
    \label{eq:weight-emission}
    w_{\rm ev}\rightarrow w_{\rm ev}\times r.
  \end{equation}
\item If the emission is rejected, we instead update the event weight according to
  \begin{equation}
    \label{eq:weight-no-emission}
    w_{\rm ev}\rightarrow w_{\rm ev}\times (2-r).
  \end{equation}
\end{itemize}
In this way we reweight both the emission probability to get the correct NLO
splitting, and the no-emission probability to get the correct NLO
Sudakov form factor.

\section{Detailed comparison of forward and backward evolution  \label{P8evol} }
\begin{figure}[htbp]
\begin{center}
\includegraphics[width=0.45\linewidth]{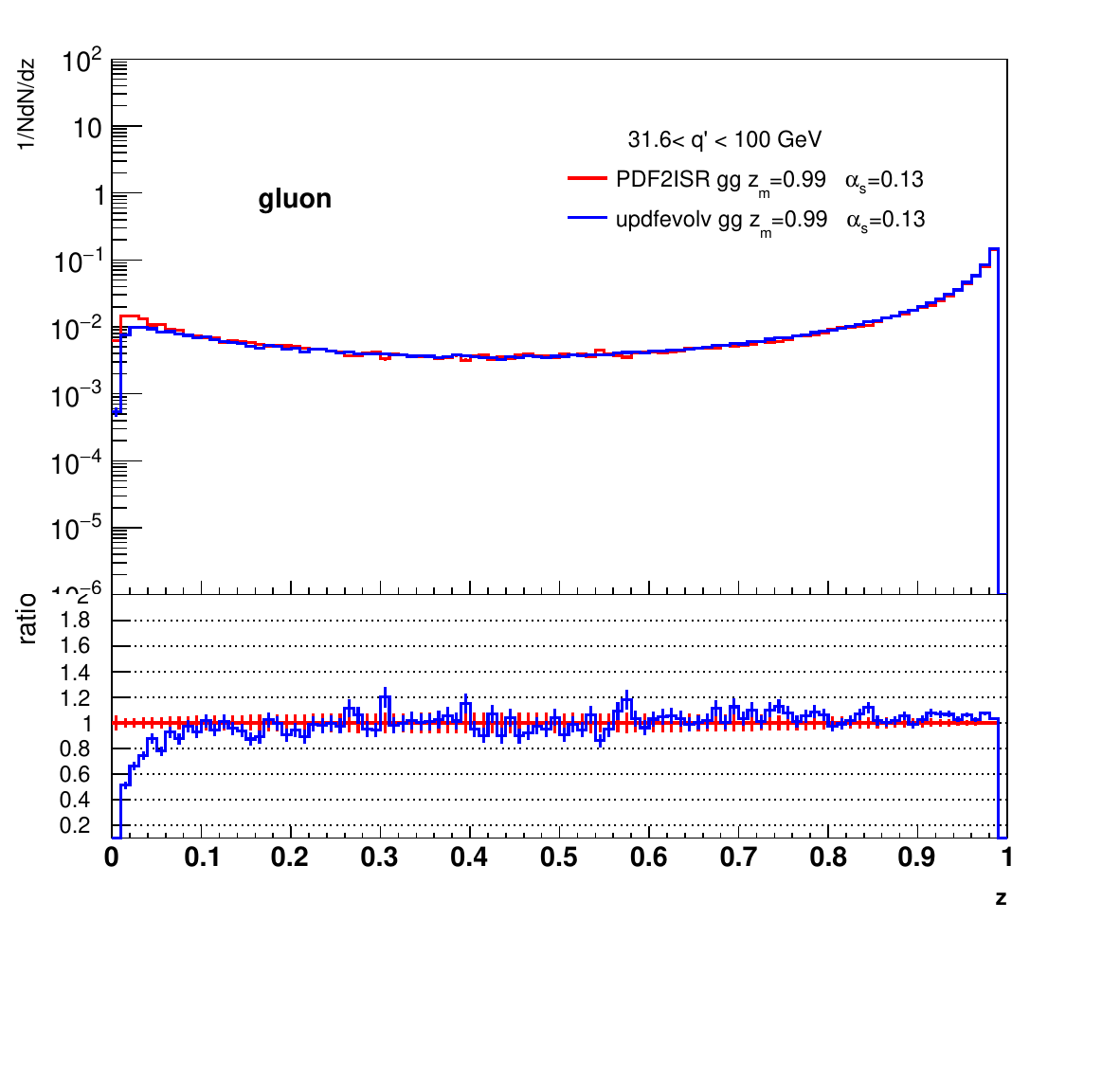} 
\includegraphics[width=0.45\linewidth]{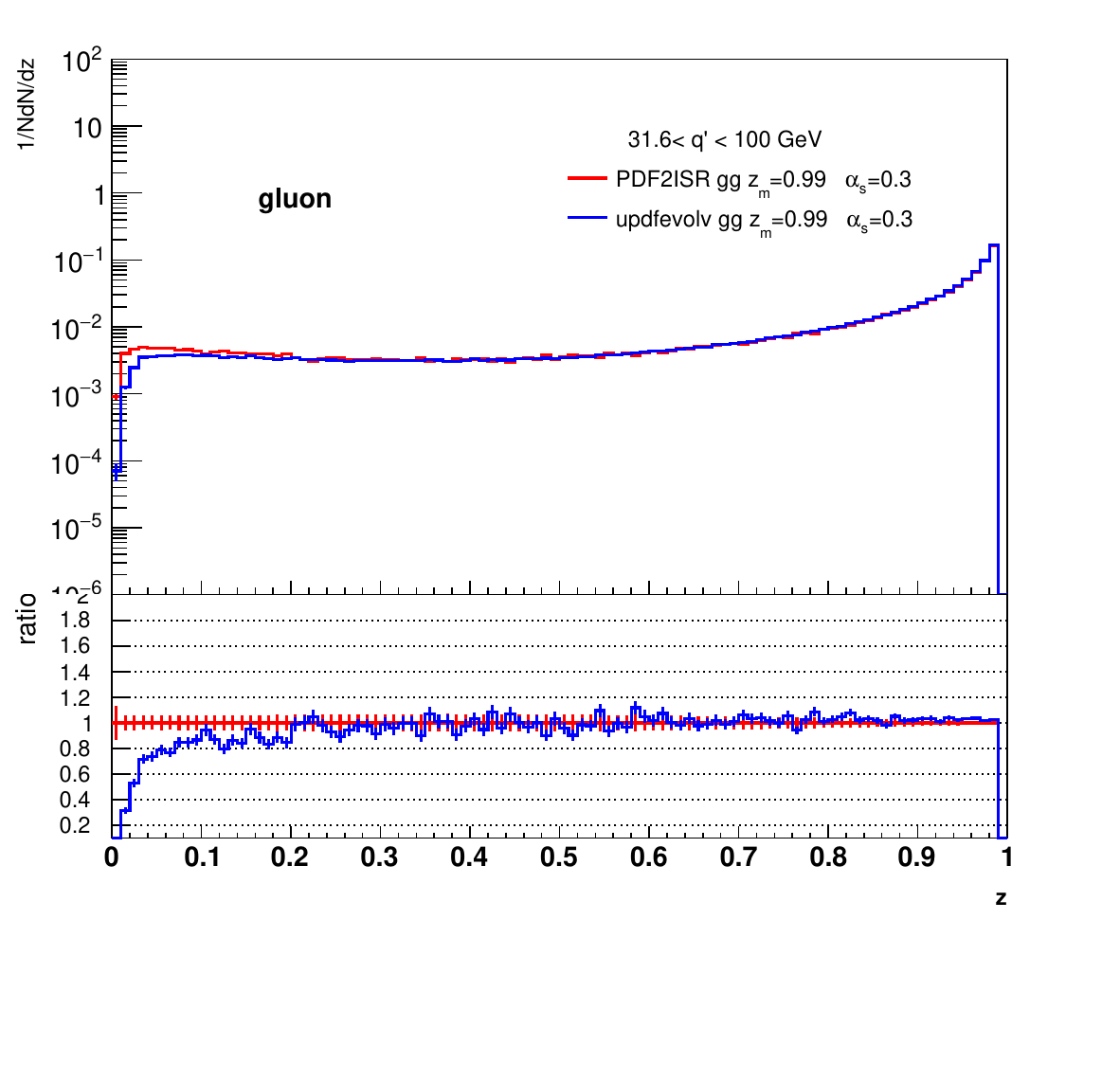} 
\vskip -1cm
\caption{Distribution of $z$ for  $g \to gg$   for $\zM=0.99$ and  $31.6 < q^\prime < 100 $~\GeV. The lower panel shows the ratio of the predictions from \pythia\  with the one from \updfevolv . Left: fixed $\as=0.13$. Right:  fixed $\as=0.3$.}
\label{Compz}
\end{center}
\end{figure}

\begin{figure}[htbp]
\begin{center}
\includegraphics[width=0.45\linewidth]{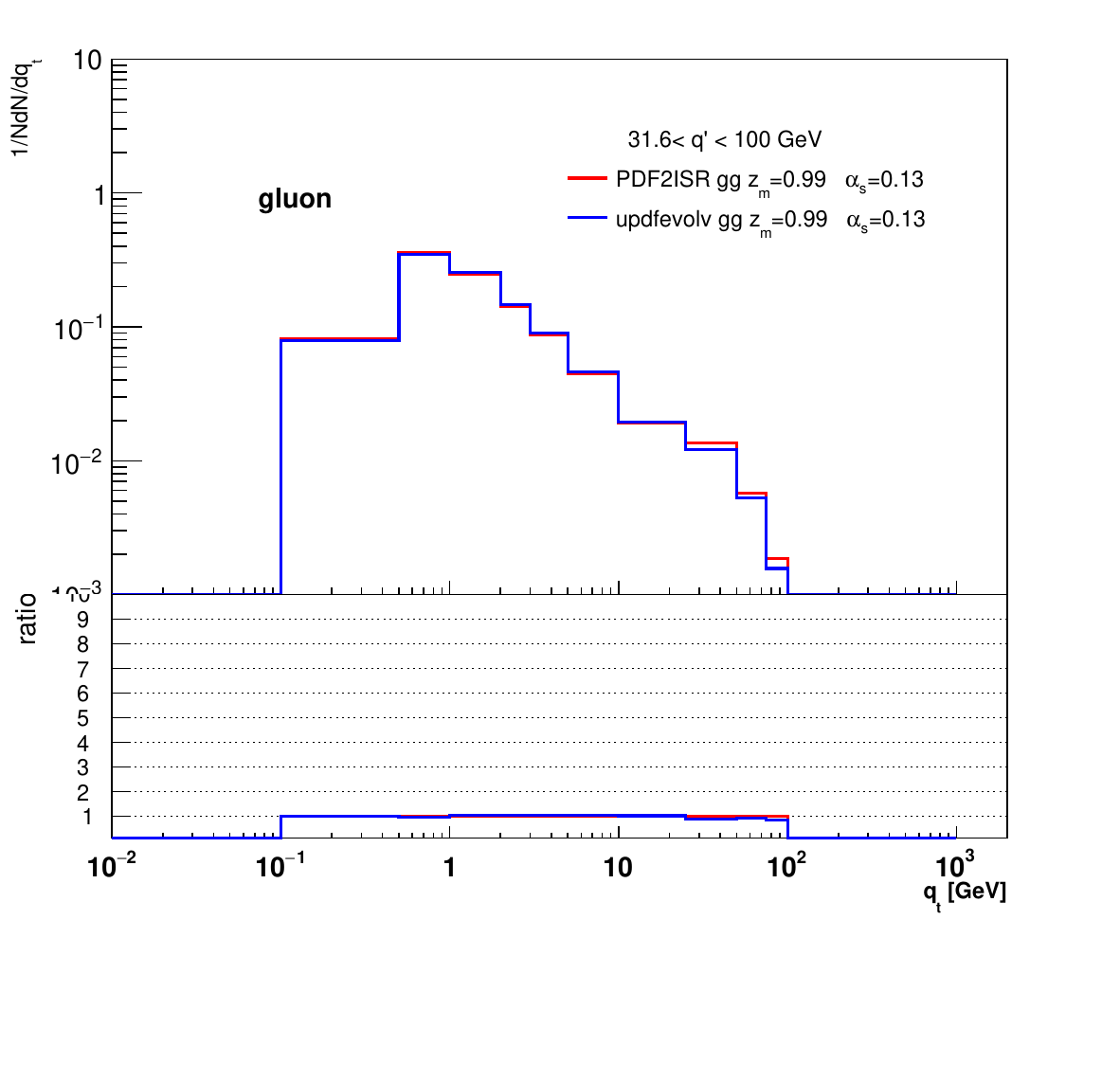} 
\includegraphics[width=0.45\linewidth]{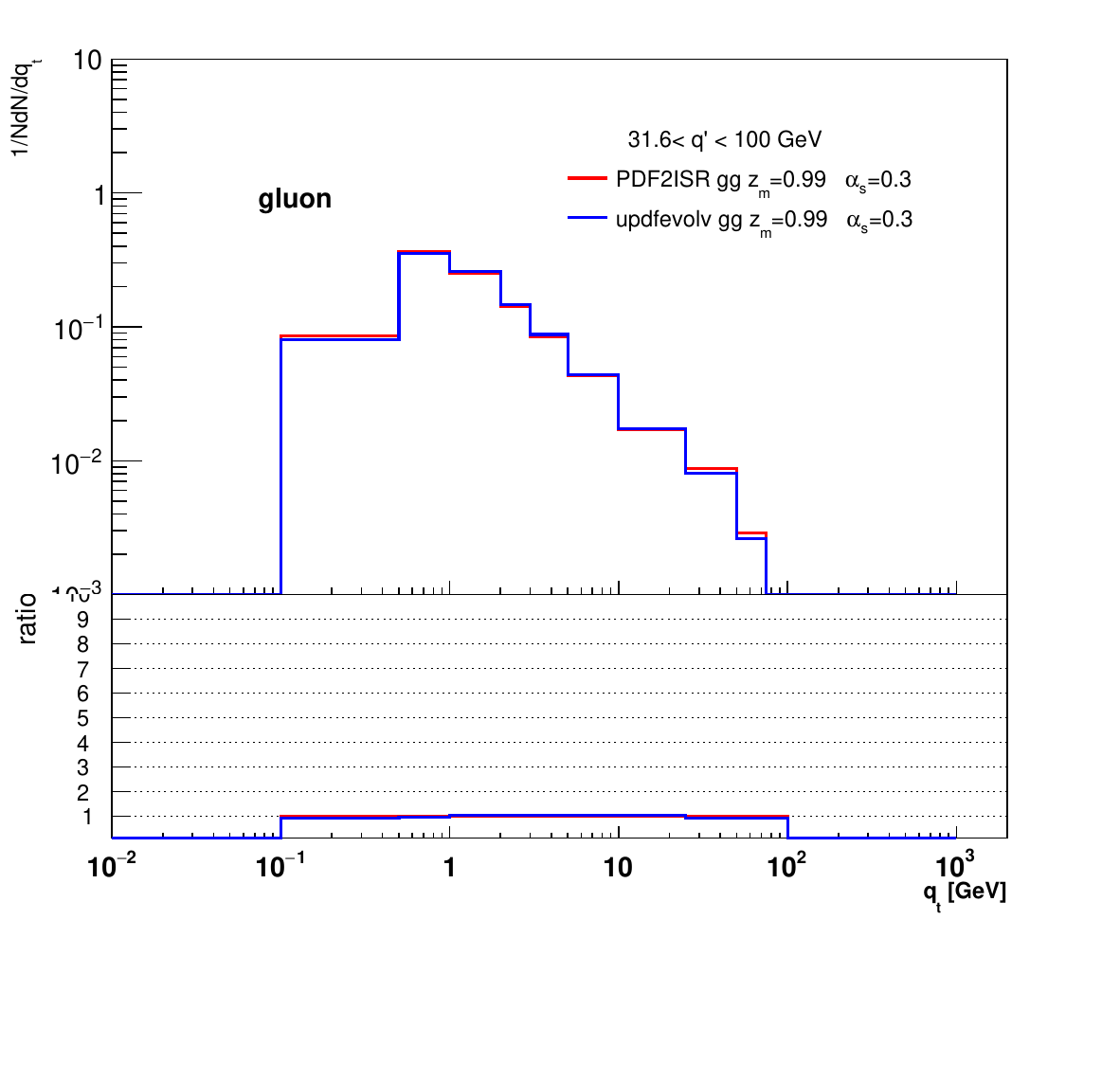} 
\caption{Distribution of \qt for  $g \to gg$   for $\zM=0.99$ and $31.6 < q^\prime < 100 $~\GeV. The lower panel shows the ratio of the predictions from \pythia\  with the one from \updfevolv . Left:   fixed $\as=0.13$. Right:  fixed $\as=0.3$}
\label{Compqt}
\end{center}
\end{figure}

In the following, we perform a detailed comparison of the forward evolution, as used in \updfevolv, with the backward evolution implemented in the initial-state parton shower of \pythia\ with the \pythiaPB\ modifications described above. A simplified scenario is used, where only $g \to gg$ splittings are considered with the LO splitting function, a fixed cutoff $\zM = 0.99$ and fixed \as\ at $\as = 0.13 \,(0.3)$. 
In the evolution, the  scale $q^\prime$ is generated from the Sudakov form-factor  $\Delta_{bw}$ (eq.(\ref{Suda})),  and the splitting variable is generated from the splitting function. The transverse momentum \qt\ of the emitted parton (see Fig.~\ref{fig:kine}) is then calculated (assuming angular ordering) via $\qt = q^\prime  (1 -z) $.
In Fig.~\ref{Compz} the distribution of the splitting variable $z$ is shown for a small slice of evolution scales $31.6 < q^\prime < 100 $~\GeV .  In Fig.~\ref{Compqt} a comparison of \qt\ is shown.
A rather good agreement between forward and backward evolution is observed.

The transverse momentum \kt\ (see Fig.~\ref{fig:kine}) is calculated from \qt : in the forward evolution in  \updfevolv\  all calculations are performed in the overall center-of-mass frame, and the final \kt\ at the end of the evolution is  given by~\cite{Hautmann:2017fcj,Jung:2024uwc}
\begin{equation}
{\bf k} = {\bf k}_0- \sum_i {\bf q}_{t,i} \ . 
\end{equation}
where ${\bf k}_0$ comes from the intrinsic \kt -distribution (which is neglected here).

\begin{figure}[htbp]
\begin{center}
\includegraphics[width=0.45\linewidth]{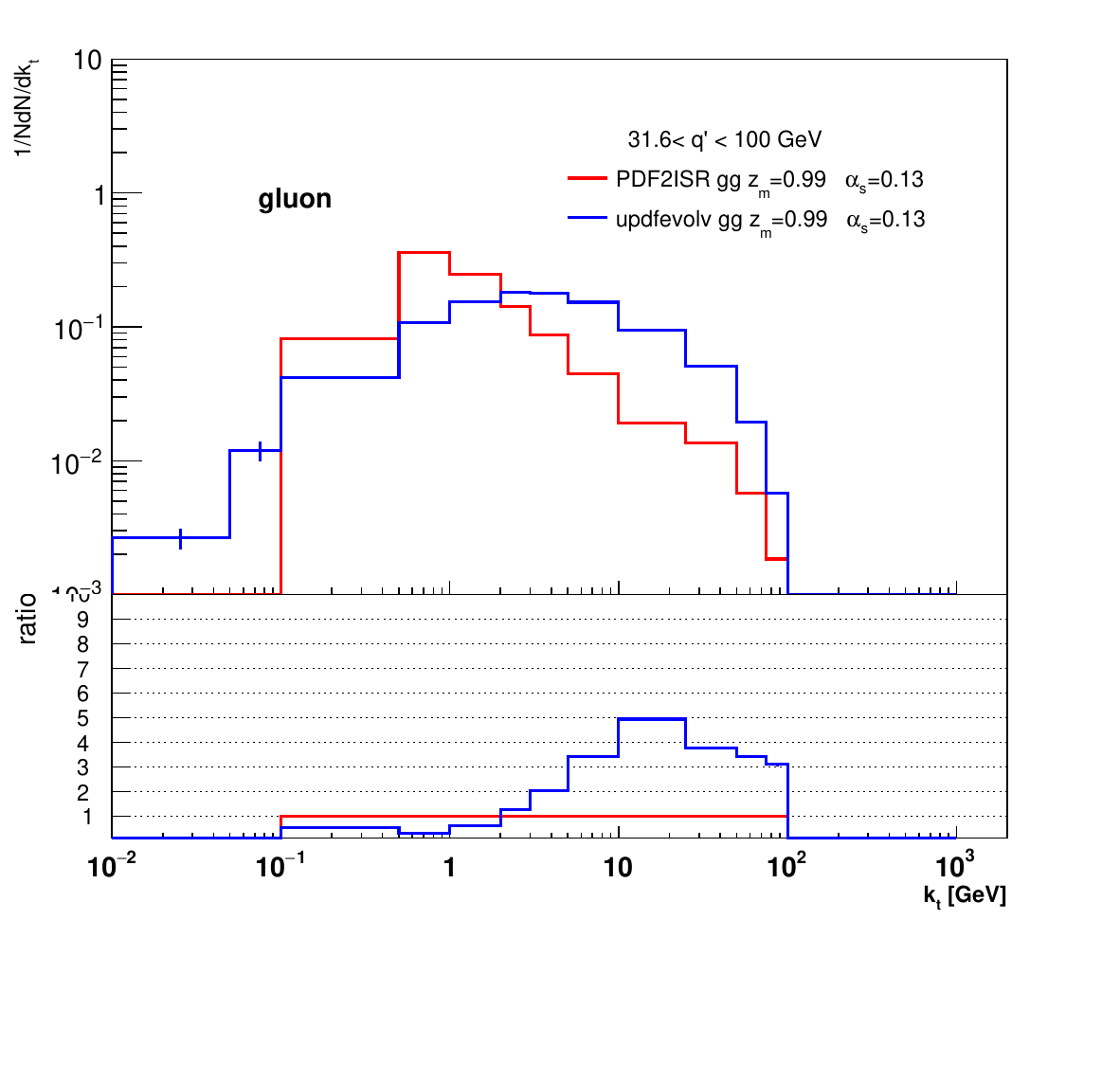} 
\includegraphics[width=0.45\linewidth]{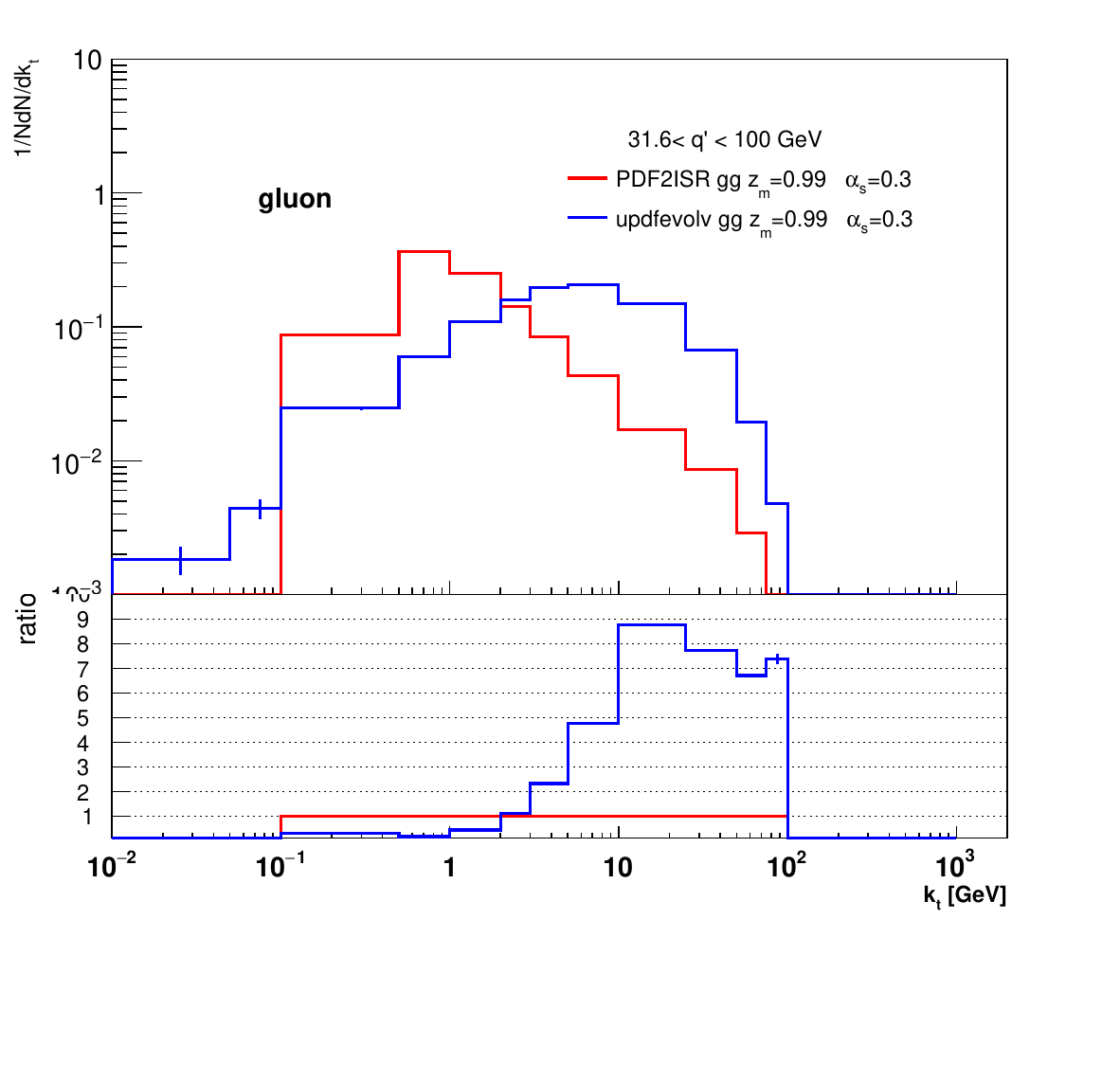} 
\caption{Distribution of \kt for  $g \to gg$   for $\zM=0.99$. The lower panel shows the ratio of the predictions from \pythia\  with the one from \updfevolv . Left:  fixed $\as=0.13$. Right:  fixed $\as=0.3$}
\label{Compkt}
\end{center}
\end{figure}

In the backward evolution in \pythia, the transverse momentum \qt\ is defined in the collinear parton-parton center-of-mass frame, and the configuration is then boosted to the overall center-of-mass frame. 
The forward and backward evolution differ  in the frame in which \qt\ is defined, and therefore differences  are expected for \kt . In the case of large \as , more emissions appear, and therefore a larger difference is expected.
A comparison of the distributions of \kt\ for different \as -values is shown in Fig.~\ref{Compkt}. One observes differences (significantly larger than for the \qt -distributions). The differences between backward and forward evolution are also significantly larger for $\as=0.3$ compared to the case with $\as=0.13$.

The differences observed in this simplified case help to explain the differences (especially in the gluon channel) observed for the Set2 scenario, where due to \as (\kt), rather large values of \as\ can be reached for small \kt\ (similar to the example above with fixed $\as=0.3$).

\begin{figure}[htbp]
\begin{center}
\includegraphics[width=0.45\linewidth]{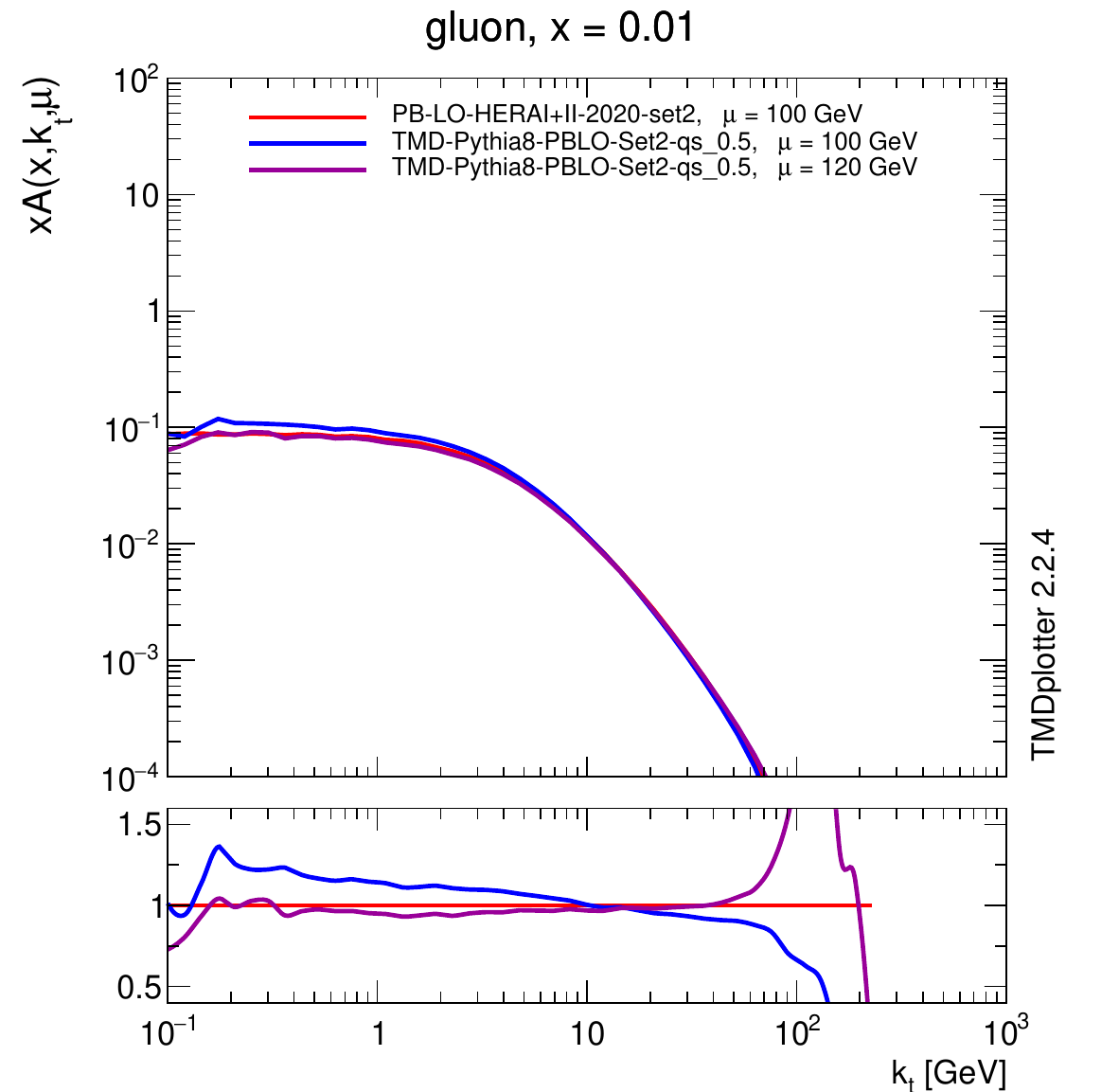} 
\caption{\small Transverse momentum distributions for gluons at $\mu=100 $~\GeV . The red line is from \updfevolv , the blue line is obtained from  \pythia\  \pythiaPB\ with LO splitting functions. The purple line shows the result with a slightly shifted scale $\mu=120 $~\GeV .  }
\label{CompScale}
\end{center}
\end{figure}

The differences in the \kt -distribution, coming from the frame in which \qt\ is defined, can be associated as a systematic uncertainty, which, however, is covered already by a scale uncertainty of the TMD distribution. In Fig.~\ref{CompScale}, we show a comparison of the \kt -distribution obtained with  the full forward evolution in \updfevolv\ and  the backward evolution in \pythia . In addition is shown the prediction with a slightly shifted scale $\mu$, showing that the differences are covered by a small variation of the scale $\mu$.

\section{Parameter settings in \pythia. \label{P8param} }
Here, we give a full list of the parameters used in \pythia. Note that
setting \texttt{SpaceShower:PB} and the \texttt{Tune:pp} codes 100001
-- 100004 are only available in our modifications of the code.

\begin{footnotesize}
\vspace{0.5cm}
\begin{minipage}{0.45\textwidth}
\subsection*{Settings for \PBNLOset\ Set1}
\vfill
\begin{verbatim}
Tune:pp =                   100001
SpaceShower:PB =            1
PDF:pSet =
  LHAPDF6:PB-TMDNLO-HERAI+II-2018-set1
BeamRemnants:primordialKThard = 0.5
SpaceShower:alphaSvalue =   0.118
SpaceShower:alphaSorder =   2
SpaceShower:pT0Ref =        0.0
SpaceShower:pTmin =         1.38
SpaceShower:pTmaxFudge=     1.0
SpaceShower:rapidityOrder = false
SpaceShower:MEcorrections = false
SpaceShower:samePTasMPI =   false
1:m0 =                      0.
2:m0 =                      0.
3:m0 =                      0.
4:m0 =                      1.47
5:m0 =                      4.5
6:m0 =                      173.
\end{verbatim}
\vfill
\end{minipage}%
\hspace{1.cm}
\begin{minipage}{0.45\textwidth}
\subsection*{Settings for \PBNLOset\ Set2}
\vfill
\begin{verbatim}
Tune:pp =                   100002
SpaceShower:PB =            2
PDF:pSet =
  LHAPDF6:PB-TMDNLO-HERAI+II-2018-set2
BeamRemnants:primordialKThard = 0.5
SpaceShower:alphaSvalue =   0.118
SpaceShower:alphaSorder =   2
SpaceShower:pT0Ref =        0.0
SpaceShower:pTmin =         1.18
SpaceShower:pTmaxFudge =    1.0
SpaceShower:rapidityOrder = false
SpaceShower:MEcorrections = false
SpaceShower:samePTasMPI =   false
1:m0 =                      0.
2:m0 =                      0.
3:m0 =                      0.
4:m0 =                      1.47
5:m0 =                      4.5
6:m0 =                      173.
\end{verbatim}
\vfill
\end{minipage}

\begin{minipage}{0.45\textwidth}
\subsection*{Settings for \PBset\ Set1}
\vfill
\begin{verbatim}
Tune:pp =                   100003
SpaceShower:PB =            1
PDF:pSet =
  LHAPDF6:PB-TMDLO-HERAI+II-2018-set1
BeamRemnants:primordialKThard = 0.5
SpaceShower:alphaSvalue =   0.13
SpaceShower:alphaSorder =   1
SpaceShower:pT0Ref =        0.0
SpaceShower:pTmin =         1.38
SpaceShower:pTmaxFudge =    1.0
SpaceShower:rapidityOrder = false
SpaceShower:MEcorrections = false
SpaceShower:samePTasMPI =   false
1:m0 =                      0.
2:m0 =                      0.
3:m0 =                      0.
4:m0 =                      1.47
5:m0 =                      4.5
6:m0 =                      173.
\end{verbatim}
\vfill
\end{minipage}%
\hspace{1.cm}
\begin{minipage}{0.45\textwidth}
\subsection*{Settings for \PBset\ Set2}
\vfill
\begin{verbatim}
Tune:pp =                   100004
SpaceShower:PB =            2
PDF:pSet = 
  LHAPDF6:PB-TMDLO-HERAI+II-2018-set2
BeamRemnants:primordialKThard = 0.5
SpaceShower:alphaSvalue =   0.13
SpaceShower:alphaSorder =   2
SpaceShower:pT0Ref =        0.0
SpaceShower:pTmin =         1.38
SpaceShower:pTmaxFudge =    1.0
SpaceShower:rapidityOrder = false
SpaceShower:MEcorrections = false
SpaceShower:samePTasMPI   = false
1:m0 =                      0.
2:m0 =                      0.
3:m0 =                      0.
4:m0 =                      1.47
5:m0 =                      4.5
6:m0 =                      173.
\end{verbatim}
\vfill
\end{minipage}
\end{footnotesize}

\section{TMDs for  \PBset\ \label{TMDLO} }
The LO \PBM\ collinear and TMD sets \PBset\ were obtained in Ref.~\cite{PBLO}, applying a starting scale $\mu_0=1.38$~\GeV and $\asmz = 0.13$ (at LO). 

\begin{figure}[t]
\centering
\includegraphics[width=0.35\linewidth]{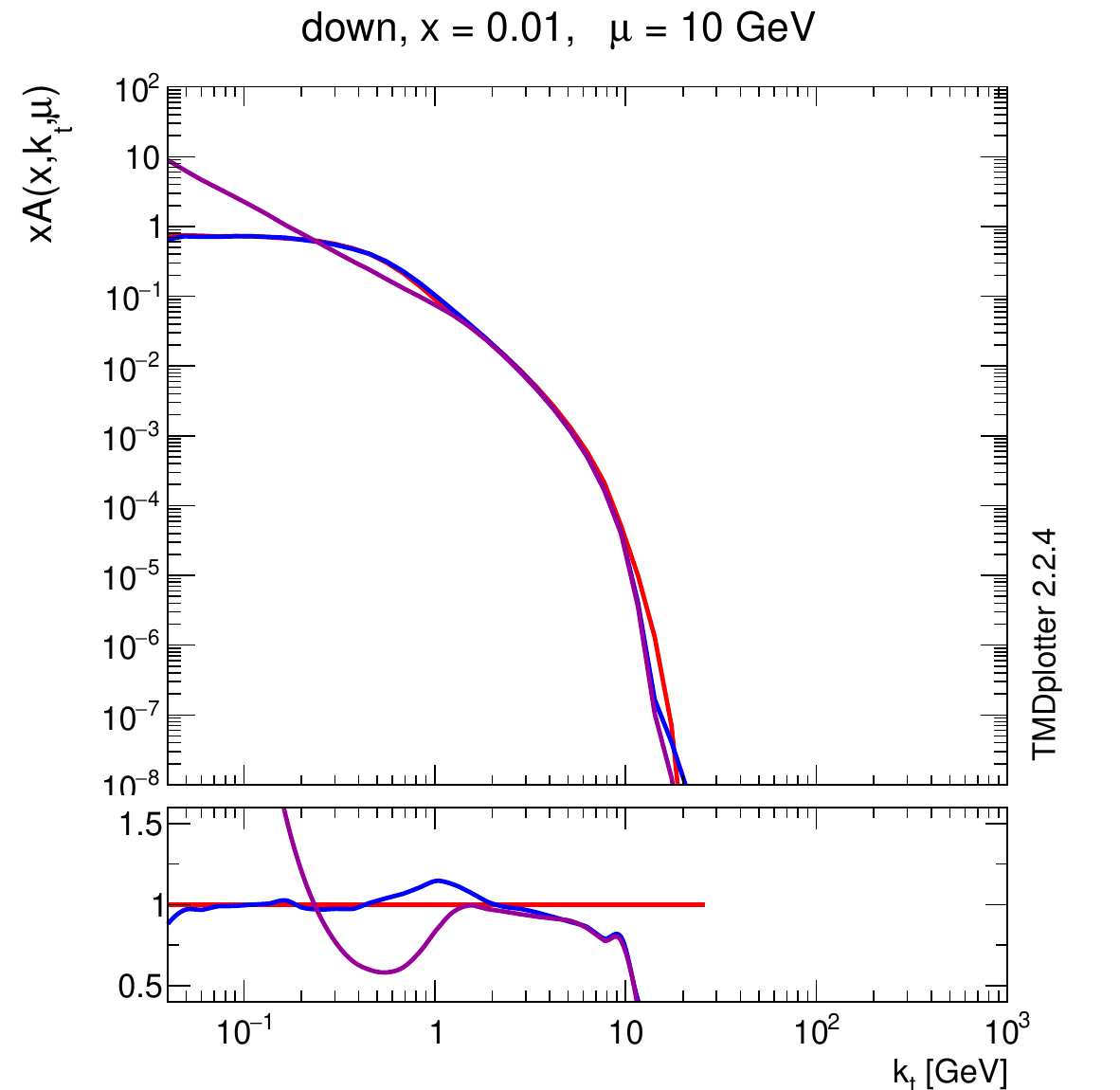} 
\includegraphics[width=0.35\linewidth]{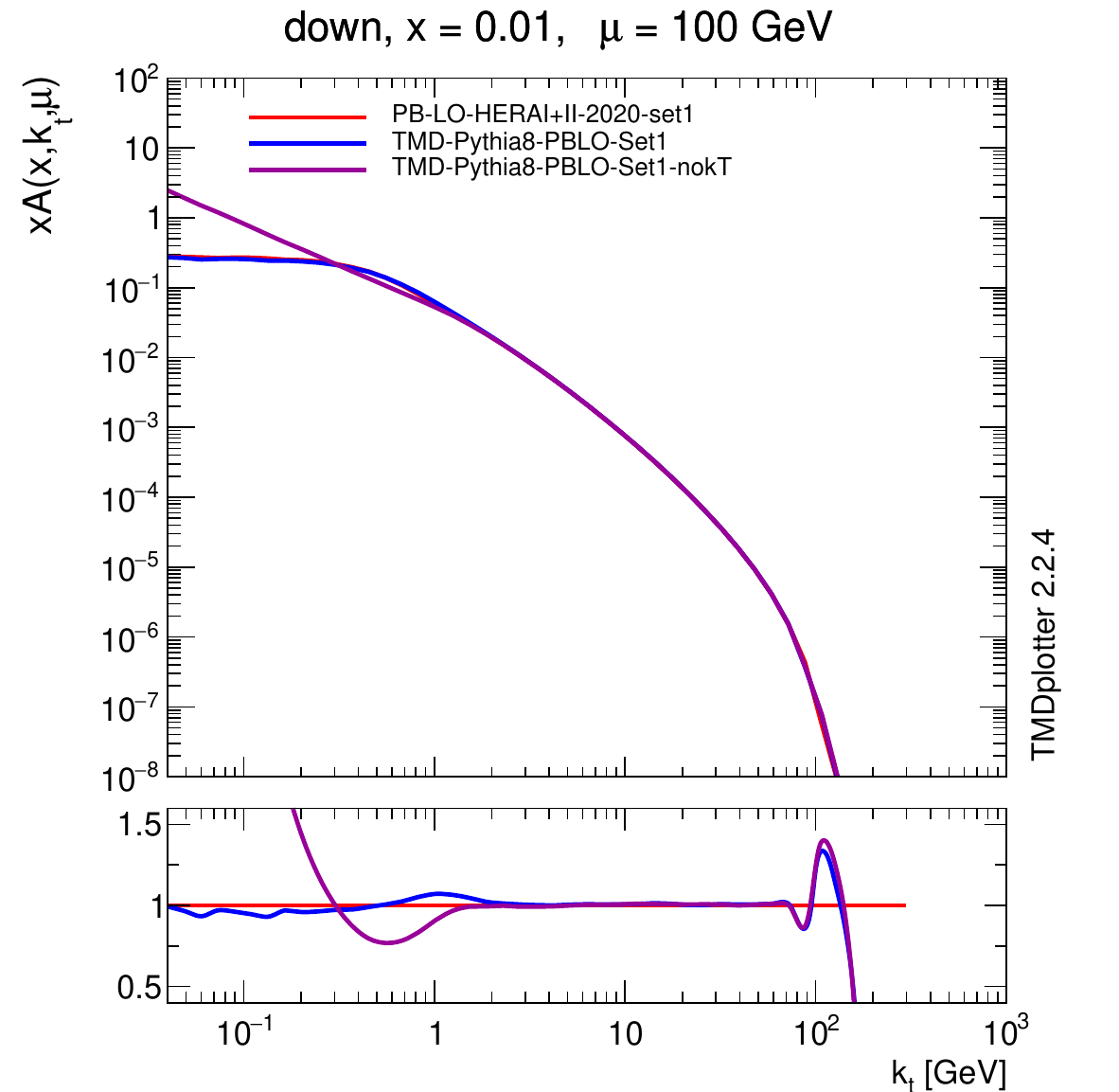}  
\includegraphics[width=0.35\linewidth]{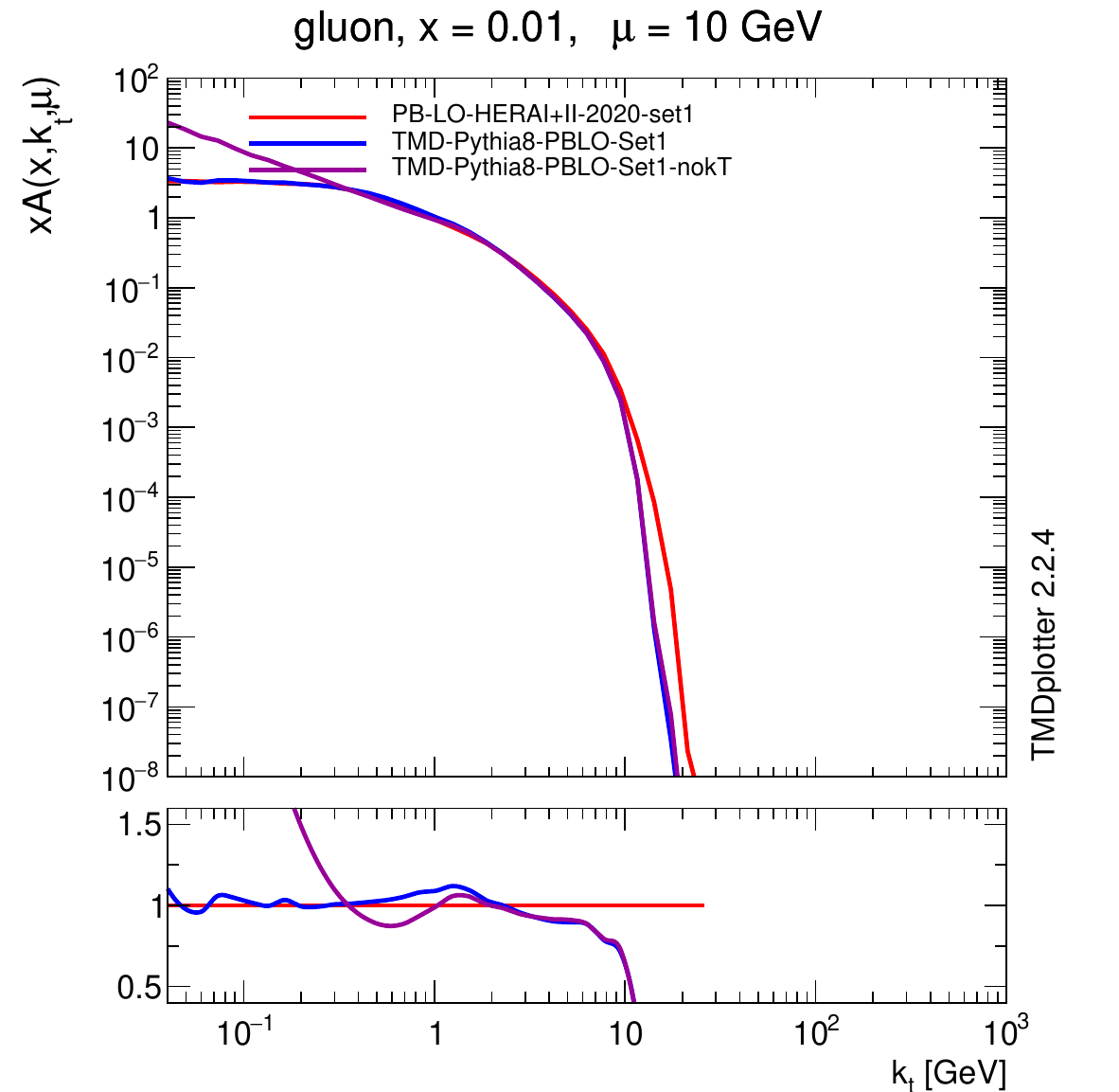} 
\includegraphics[width=0.35\linewidth]{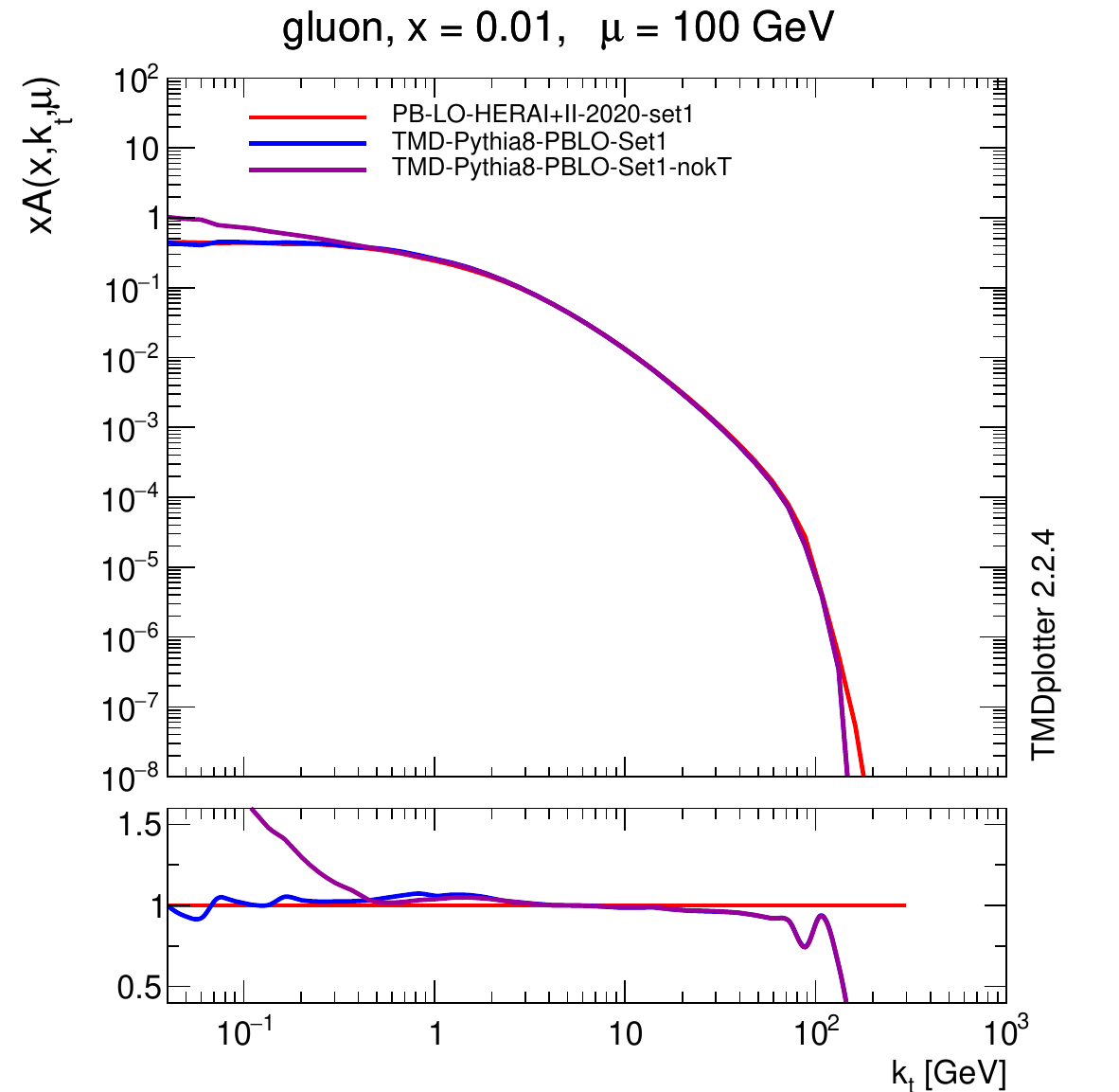}  
\cprotect\caption {\small Transverse momentum distributions for gluons and down-quarks as $\mu=10 (100) $\GeV , obtained from \PBset~Set1 evolved from a starting scale $\mu_0=1.38$ \GeV\ and \protect\pythia\ \pythiaPB applying  \verb+pTmin=1.38+ \GeV .
The \PBset~Set1 predictions are obtained at LO (with LO $\asmz = 0.130$).  
\label{fig:PB-P8Set1-LO-noKT-AngOrd} }
\end{figure}

In Fig.~\ref{fig:PB-P8Set1-LO-noKT-AngOrd}, the down quark and gluon distributions at LO obtained with \pythia\  \pythiaPB\ applying  \verb+pTmin=1.38+~\GeV\  are shown and compared  with those obtained with the LO \PBM - TMD distributions for Set1 conditions \PBset~Set1. A very good agreement is observed between the \PBM forward evolution and the \pythia\ parton shower in a backward evolution.
The prediction without including intrinsic \kt\ is also shown for comparison.

\begin{figure}[t]
\centering
\includegraphics[width=0.35\linewidth]{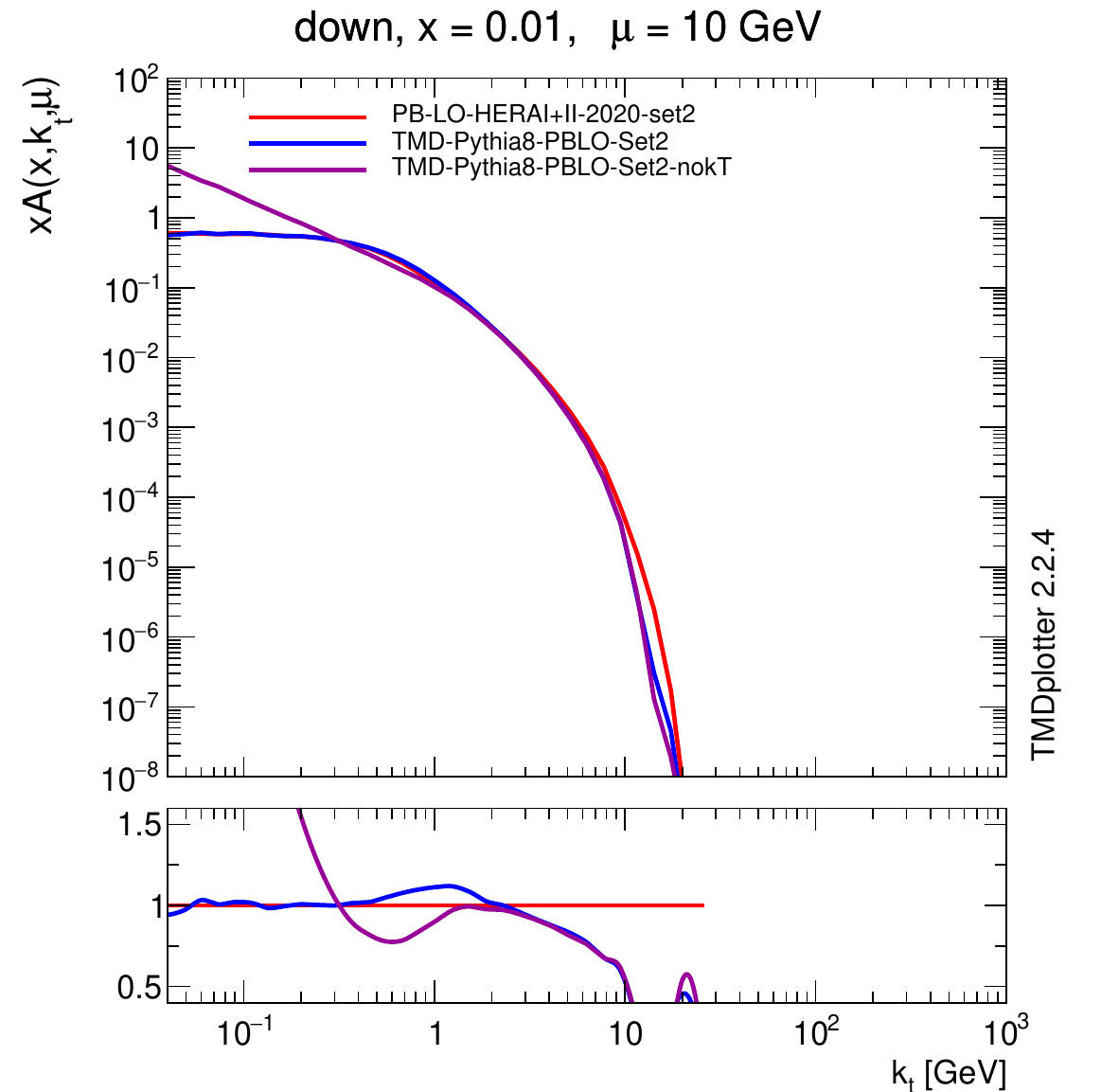} 
\includegraphics[width=0.35\linewidth]{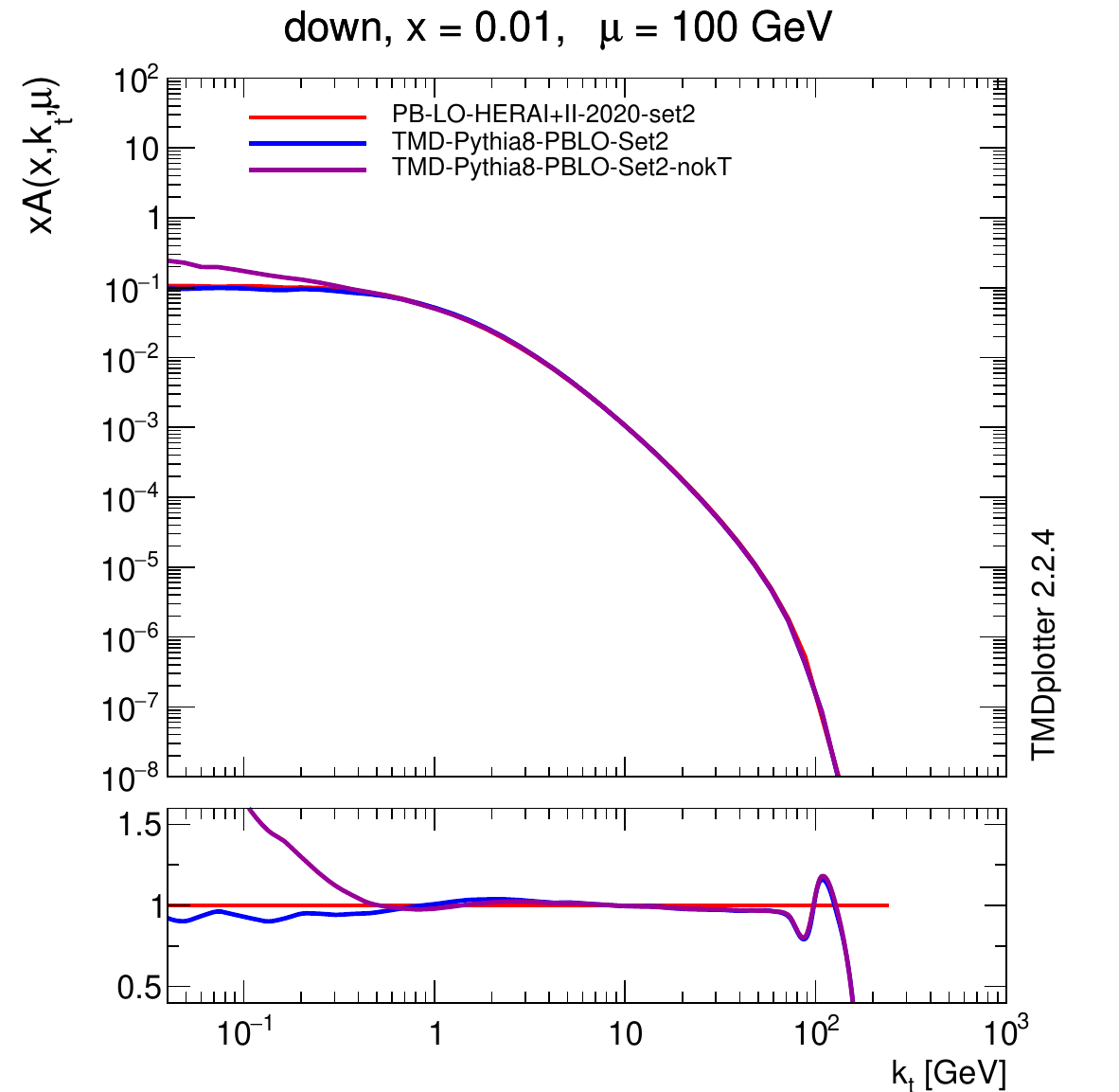}  
\includegraphics[width=0.35\linewidth]{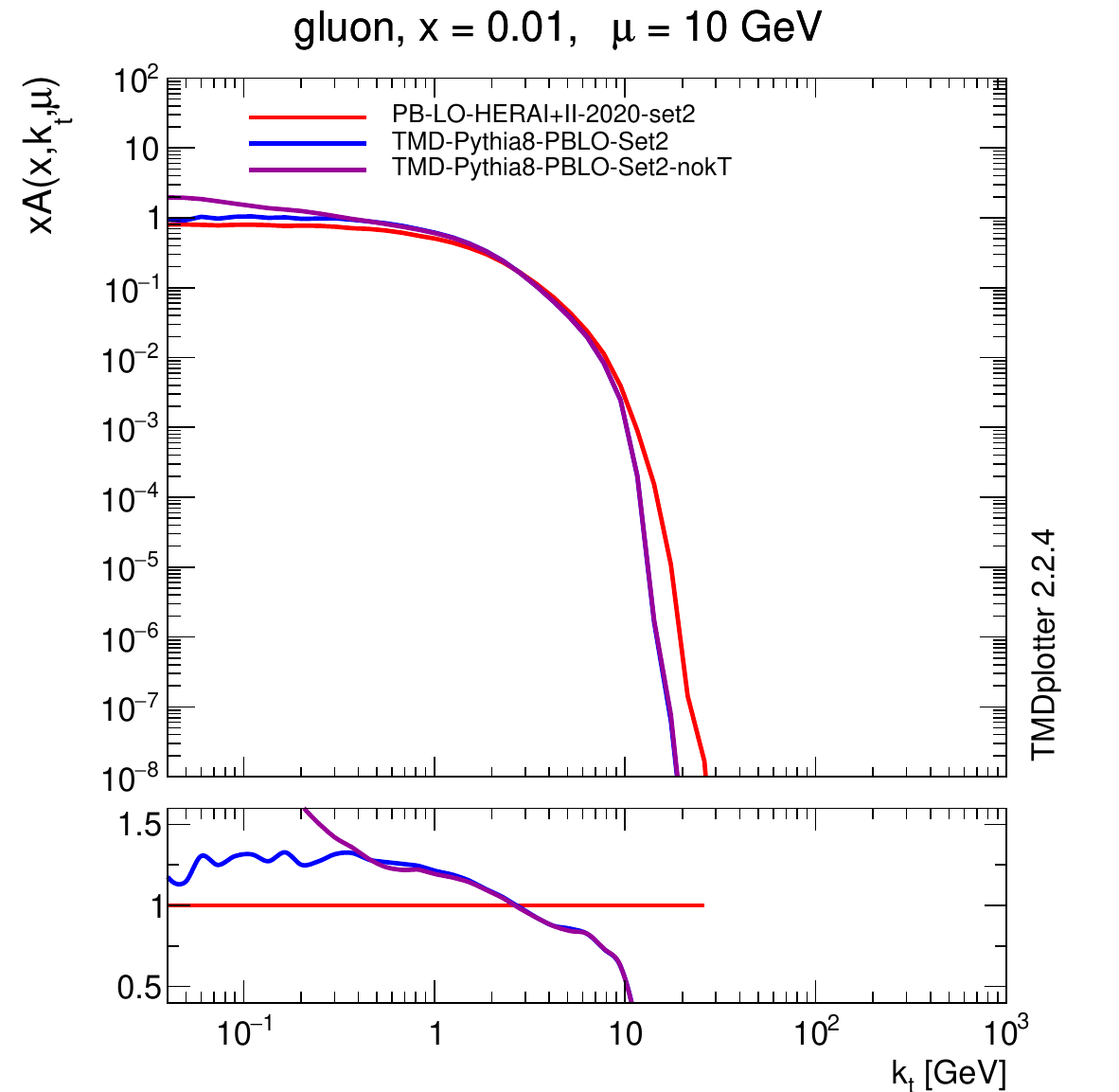} 
\includegraphics[width=0.35\linewidth]{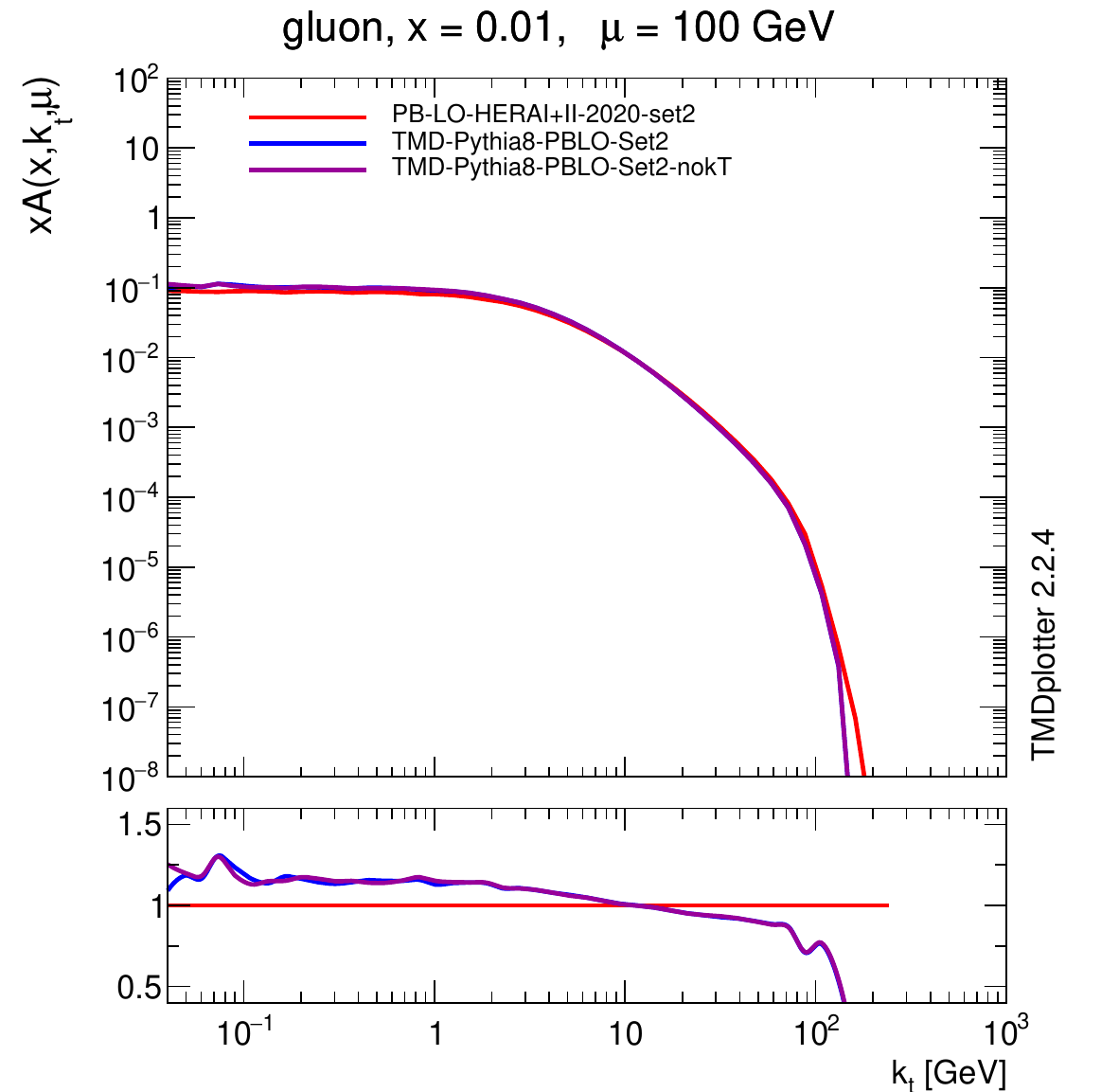}  
\cprotect\caption {\small Transverse momentum distributions for gluons and down-quarks as $\mu=10~ (100) $ \GeV , obtained from \PBset~Set2 evolved from a starting scale $\mu_0=1.38$ \GeV\ and \protect\pythia\ \pythiaPB applying  \verb+pTmin=1.38+ \GeV .
The \PBset~Set2 predictions are obtained at LO (with LO $\asmz = 0.130$).  
\label{fig:PB-P8Set2-LO-noKT-AngOrd} }
\end{figure}

A comparison of quark and gluon distributions of \PBset~Set2 obtained within the \PBM -approach with those from \pythia\  \pythiaPB after applying the appropriate scale change in \as\ is shown in Fig.~\ref{fig:PB-P8Set2-LO-noKT-AngOrd} for different scales $\mu$. Once again, a strong and satisfactory agreement is observed for the quark distribution. The comparison of the gluon distribution is affected by the different reference frames as explained in Appendix~\ref{P8evol}.

\bibliographystyle{mybibstyle-new.bst}
\raggedright  
\bibliography{/Users/jung/Bib/hannes-bib}

\end{document}